\documentclass[10pt,twocolumn,letterpaper]{article}
\usepackage{usenix, times,color,graphicx,amsmath,array,booktabs, amssymb, subcaption}
\usepackage{pifont}
\usepackage{xspace}
\usepackage{color}
\usepackage{xcolor}
\usepackage{adjustbox}
\usepackage{multirow}
\usepackage{colortbl}
\usepackage{comment}
\usepackage[small,compact]{titlesec}
\usepackage{wrapfig}
\usepackage[
  colorlinks=true,
  citecolor=red
]{hyperref}
\AtBeginDocument{%
  \providecommand\BibTeX{{%
    \normalfont B\kern-0.5em{\scshape i\kern-0.25em b}\kern-0.8em\TeX}}}

\newcommand\blfootnote[1]{%
  \begingroup
  \renewcommand\thefootnote{}\footnote{#1}%
  \addtocounter{footnote}{-1}%
  \endgroup
}

\renewcommand{\baselinestretch}{1.0}

\begin{document}

\title{\tool: Isolated and Adaptive Swapping for Multi-Applications \\on Remote Memory}

\author{\rm{Chenxi Wang}$^{\dag\ast}$\hspace{1.2em}
Yifan Qiao$^{\dag\ast}$\hspace{1.2em}
Haoran Ma$^{\dag}$\hspace{1.2em}
Shi Liu$^{\dag}$\hspace{1.2em}
Yiying Zhang$^{\ddag}$\\[.3em]
\rm{Wenguang Chen}$^{\S}$\hspace{1.2em}
Ravi Netravali$^{\sharp}$\hspace{1.2em}
Miryung Kim$^{\dag}$\hspace{1.2em}
Guoqing Harry Xu$^{\dag}$
\\[.3em]
UCLA$^{\dag}$\hspace{1.2em}
UCSD$^{\ddag}$\hspace{1.2em}
Tsinghua University$^\S$\hspace{1.2em}
Princeton University$^{\sharp}$
}

\newcommand{\cf}{\hbox{\emph{cf.}}\xspace}
\newcommand{\etal}{\hbox{\emph{et al.}}\xspace}
\newcommand{\eg}{\hbox{\emph{e.g.}}\xspace}
\newcommand{\ie}{\hbox{\emph{i.e.}}\xspace}
\newcommand{\st}{\hbox{\emph{s.t.}}\xspace}
\newcommand{\wrt}{\hbox{\emph{w.r.t.}}\xspace}
\newcommand{\etc}{\hbox{\emph{etc.}}\xspace}
\newcommand{\viz}{\hbox{\emph{viz.}}\xspace}

\newcommand{\squishlist}{
   \begin{list}{$\bullet$}
    { \setlength{\itemsep}{0pt}      \setlength{\parsep}{3pt}
      \setlength{\topsep}{3pt}       \setlength{\partopsep}{0pt}
      \setlength{\leftmargin}{1.0em} \setlength{\labelwidth}{1em}
      \setlength{\labelsep}{0.5em} } }
\newcommand{\squishend}{
    \end{list}  }

\newcommand{\mysection}[1]{\section{#1}}
\newcommand{\mysubsection}[1]{\subsection{#1}}
\newcommand{\mysubsubsection}[1]{\vspace{-.3em}\subsubsection{#1}}
\definecolor{ForestGreen}{RGB}{34,139,34}
\newcommand{\cw}[1] {{\textcolor{ForestGreen}{Chenxi: {#1}}}}
\newcommand{\hx}[1] {{\textcolor{blue}{HX: {#1}}}}
\newcommand{\rev}[1] {{\textcolor{blue}{{#1}}}}
\newcommand{\yq}[1] {{\textcolor{purple}{Yifan: {#1}}}}
\newcommand{\rn}[1] {{\textcolor{magenta}{RN: {#1}}}}
\newcommand{\miryung}[1] {{\textcolor{magenta}{Miryung: {#1}}}}
\newcommand{\MyPara}[1]{\vspace{.1em}\noindent\textit{\textbf{#1}}}

\newcommand{\tool}[0]{Canvas\xspace}
\newcommand{\us}{\emph{$\mu$s}\xspace}
\newcommand{\naive}[0]{na\"{i}ve\xspace}
\newcommand{\Naive}[0]{Na\"{i}ve\xspace}
\newcommand{\naively}[0]{na\"{i}vely\xspace}
\newcommand{\Naively}[0]{Na\"{i}vely\xspace}
\newcommand{\codeIn}[1]{{\small\texttt{#1}}}
\setcounter{page}{1}

\maketitle
\section*{Abstract}
\blfootnote{$\ast$ Contributed equally.}
Remote memory techniques for datacenter applications have recently gained a great deal of popularity. Existing remote memory techniques focus on the efficiency of a single application setting only. However, when multiple applications co-run on a remote-memory system, significant interference could occur, resulting in unexpected slowdowns even if the same amounts of physical resources are granted to each application. This slowdown stems from massive sharing in applications' swap data paths. \tool is a redesigned swap system that fully isolates swap paths for remote-memory applications. \tool allows each application to possess its dedicated swap partition, swap cache, prefetcher, and RDMA bandwidth. Swap isolation lays a foundation for adaptive optimization techniques based on each application's own access patterns and needs. We develop three such techniques: (1) adaptive swap entry allocation, (2) semantics-aware prefetching, and (3) two-dimensional RDMA scheduling. A thorough evaluation with a set of widely-deployed applications demonstrates that \tool minimizes performance variation and dramatically reduces performance degradation.

\mysection{Introduction}

Techniques enabling datacenter applications to use far memory~\cite{resource-disaggregation-osdi16,infiniswap-nsdi17,fastswap-eurosys20,lagar-cavilla-asplos19,leap-atc20, legoos-osdi18, semeru@osdi2020,aifm@osdi2020,disaggregated-runtime-asplos21} have gained traction due to their potential to break servers' memory capacity wall, thereby improving performance and resource utilization. Existing far-memory techniques can be roughly classified into two categories: (1) clean-slate techniques~\cite{aifm@osdi2020,disaggregated-runtime-asplos21} that provide new primitives for developers to manage remote memory, and (2) swap-based techniques~\cite{infiniswap-nsdi17,legoos-osdi18,fastswap-eurosys20,semeru@osdi2020, nvmeof} that piggyback on existing swap mechanisms in the OS kernel. Clean-slate techniques provide greater efficiency by enabling user-space far memory accesses, while swap-based techniques offer transparency, allowing legacy code to run \emph{as is} on a far-memory system. This paper focuses on swap mechanisms as they are more practical and easier to adopt.%

A typical swap system in the OS uses a \emph{swap partition} and \emph{swap cache} for applications to swap data between memory and external storage. The swap partition is a storage-backed swap space. The swap cache is an intermediate buffer between the \emph{local memory}
and storage\textemdash it caches \emph{unmapped pages}
that were just swapped in or are about to be swapped out. Upon a page fault, the OS looks up the swap cache; a cache miss would trigger a \emph{demand swap} and a number of \emph{prefetching swaps}. Swaps are served by RDMA and all fetched pages are initially placed in the swap cache. The demand page is then mapped to a virtual page and moved out of the swap cache, completing the fault handling process. 

\MyPara{Problems.}
Current swap systems run multiple applications over shared swap resources (\ie, swap partition, RDMA, \etc). This design works for \emph{disk-based swapping} where disk access is slow\textemdash each application can allow only a tiny number of pages to be swapped to maintain an acceptable overhead. This assumption, however, no longer holds under far memory because an application can place more data in far memory than local memory and yet still be efficient, thanks to RDMA's low latency and high bandwidth. %

As such, applications have orders-of-magnitude more swap requests under far memory than disks. Millions of swap requests from different applications go through the same shared data path in a short period of time, leading to \emph{severe performance interference}. Our experiments show that, with the same amounts of CPU and local-memory resources, co-running applications leads up to a 6$\times$ slowdown, an overhead unacceptable for any real-world deployment.

\MyPara{State of the Art.}  Interference is a known problem in datacenter applications and a large body of work exists on isolation of CPU~\cite{li-ppopp09, bartolini-taco14, cherkasova-sigmetrics07}, I/O~\cite{mclock-osdi10, shue-osdi12}, network bandwidth~\cite{ballani-sigcomm11, ghodsi-nsdi11,seawall-hotcloud10,faircloud-sigcomm12,silverline-hotcloud11,eyeq-hotcloud12} and processing~\cite{iron-nsdi18}. Most of these techniques build on Linux's \codeIn{cgroup} mechanism, which 
focuses on isolation of traditional resources such as CPU and memory, \emph{not} swap resources such as remote memory usage and RDMA. 
Prior swap optimizations such as Infiniswap~\cite{infiniswap-nsdi17} and Fastswap~\cite{fastswap-eurosys20} focus on reducing remote access latency, overlooking the impact of swap interference in realistic settings. Justitia~\cite{justitia-nsdi22} isolates RDMA bandwidth between applications, but does not eliminate other types of interference such as locking and swap cache usage. 

\MyPara{Contribution \#1: Interference Study (\S\ref{sec:motivation}).}  We conducted a systematic study with a set of widely-deployed applications on Linux 5.5, 
the latest kernel version compatible with Mellanox's latest driver (4.9-3.1.5.0) for our InfiniBand card.
Our results reveal three major performance problems: 

\squishlist
\item \textbf{Severe lock contention:} Since all applications share a single swap partition, extensive locking is needed for swap entry allocation (needed by every swap-out), reducing throughput and precluding full utilization of RDMA's bandwidth. Our experience shows that in windows of frequent remote accesses, applications can spend \textbf{70\%} of the windows' time on swap entry allocation. %

\item \textbf{Uncontrolled use of swap resources (\eg, RDMA):} The use of the shared RDMA bandwidth is often dominated by the pages fetched for applications with many threads simultaneously performing frequent remote accesses. For example, aggressively (pre)fetching pages to fulfill one application's needs can disproportionally reduce other applications' bandwidth usage. 
Further, even within one application, prefetching competes for resources with demand swaps, leading to either prolonged fault handling or delayed prefetching that fails to bring back pages in time. 

\item \textbf{Reduced prefetching effectiveness:} Applications use the same prefetcher, prefetching data based on \emph{low-level (sequential or strided) access patterns} across applications. However, modern applications exhibit far more diverse access patterns, making it hard for prefetching to be effective across the board. For example, co-running Spark and native applications reduces Leap~\cite{leap-atc20}'s prefetching contribution by \textbf{3.19$\times$}.%

\squishend

These results highlight two main problems. First, interference is caused by sharing a combination of swap resources including the swap partition/cache, and RDMA (bandwidth and SRAM on RNIC). Although recent kernel versions added support~\cite{patch3} for charging prefetched pages into \codeIn{cgroup}, resolving interference requires a \emph{holistic} approach that can isolate all these resources. Furthermore, interference stems not only from resource racing, but also from fundamental limitations with the current design of the swap system. For instance, reducing interference between prefetching and demand swapping requires understanding whether a prefetching request can come back in time. If not, it should be dropped to give resources to demand requests, which are on the critical path. This, in turn, requires a redesign of the kernel's fault handling logic.

Second, cloud applications exhibit highly diverse behaviors and resource profiles. For example, applications with a great number of threads are more sensitive to locking than single-threaded applications. Furthermore, managed applications such as Spark often make heavy use of reference-based data structures while native applications are often dominated by large arrays. The \emph{application-agnostic nature} of the swap system makes it hard for a one-size-fits-all policy (\eg, a global prefetcher) to work well for diverse applications. Effective per-application policies dictates (1) holistic swap isolation and (2) understanding application semantics, which is currently inaccessible in the kernel.

\MyPara{Contribution \#2: Holistic Swap Isolation (\S\ref{sec:isolation}).} To solve the first problem, we develop \tool, a \emph{fully-isolated} swap system, which enables each application to have its dedicated swap partition, swap cache, and RDMA usage. 
In doing so, \tool can charge each application's \codeIn{cgroup} for the usage of all kinds of swap resources, preventing certain applications from aggressively invading others' resources.

\MyPara{Contribution \#3: Isolation-Enabled Adaptive Optimizations (\S\ref{sec:customization}).} 
To solve the second problem, we develop a set of adaptive optimizations that can tailor their policies and strategies to application-specific swap behaviors and resource needs. 
Our adaptive optimizations bring a \emph{further boost} on top of the isolation-provided benefits, making co-running applications even \emph{outperform} their individual runs.

\textbf{(1) Adaptive Swap Entry Allocation (\S\ref{sec:alloc})} Separating swap partitions reduces lock contention at swap entry allocations to a certain degree, but the contention can still be heavy for multi-threaded applications. For example, Spark creates many threads to fully utilize cores and these threads need synchronizations before obtaining swap entries. The synchronization overhead increases dramatically with the number of cores (\S\ref{sec:swap_optimization}), creating a scalability bottleneck. We develop an adaptive swap entry allocator that dynamically balances between the degree of lock contention (\ie, time) and the amount of swap space needed (\ie, space) based on each application's memory behaviors.

\textbf{(2) Adaptive Two-tier Prefetching (\S\ref{sec:prefetching})} 
Current kernel prefetchers build on low-level access patterns (\eg, sequential or strided). Although such patterns are useful for applications with large array usages, many cloud applications are written in high-level, managed languages such as Java or Python; their accesses come from multiple threads or exhibit pointer-chasing behavior as opposed to sequential or strided patterns.
As effective prefetching is paramount to remote-memory performance, \tool employs a two-tier prefetching design. Our \emph{kernel-tier prefetcher} prefetches data for each application into its private swap cache based on low-level patterns. Once this prefetcher cannot effectively prefetch data, \tool adaptively forwards the faulty address up to the \emph{application tier} via a modified \codeIn{userfaultfd} interface, enabling customized prefetching logic at the level of reference-based or thread-based access patterns. 

\textbf{(3) Adaptive RDMA Scheduling (\S\ref{sec:rdma})} Isolating RDMA bandwidth alone for each application is insufficient. As there could be many more {\em prefetching} requests than {\em demand swap requests}, \naively sending all to RDMA delays demand requests, increasing fault-handling latency. On the other hand, \naively delaying prefetching requests (as in FastSwap~\cite{fastswap-eurosys20}) reduces their \emph{timeliness}, making prefetched pages useless. We built a \emph{two-dimensional} RDMA scheduler, which schedules packets not only between applications but also between prefetching and demand requests for each application.

\MyPara{Results.} 
Our evaluation (\S\ref{sec:eval}) with a set of 14 widely-deployed applications (including Spark~\cite{spark}, Cassandra~\cite{cassandra}, Neo4j~\cite{neo4j}, Memcached~\cite{memcached}, XGBoost~\cite{xgboost,chen-kdd16}, Snappy~\cite{snappy}, \etc) demonstrates that \tool improves the overall application performance by up to 
\textbf{6.2$\times$} (average \textbf{3.5$\times$}) and reduces applications' performance variation (\ie, standard deviation) by \textbf{7$\times$}, from an overall of \textbf{1.72} to \textbf{0.23}. \tool improves the overall RDMA bandwidth utilization by \textbf{2.8$\times$} for co-run applications. \tool is available at \url{https://github.com/uclasystem/canvas}.

\mysection{Background\label{sec:background}}

\begin{figure}[t]
    \centering
    \includegraphics[width=0.99\linewidth]{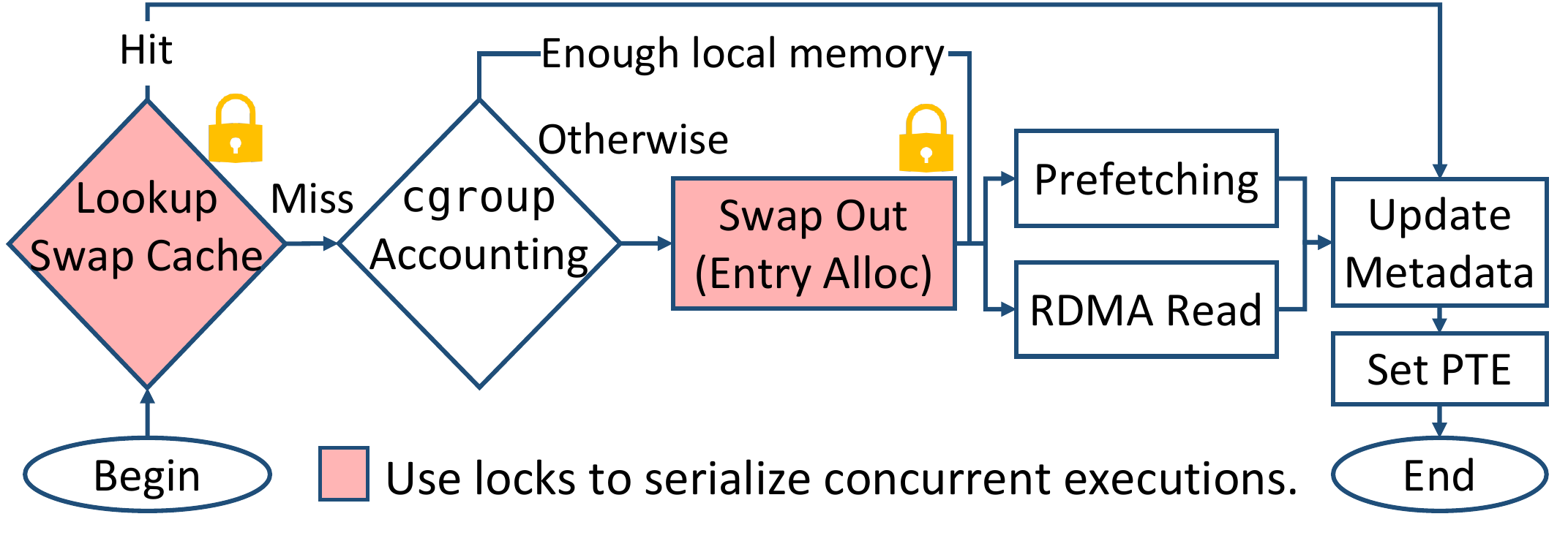}
    \vspace{-.5em}
    \caption{The kernel's remote-access data path.
    \label{fig:dataplane}}
    \vspace{-.5em}
\end{figure}

This section presents the necessary background in Linux 5.5, which is the latest kernel version compatible with Mellanox's latest driver for our InfiniBand adapter.

Figure~\ref{fig:dataplane} illustrates the kernel's remote access data path where remote memory is mapped into the host server as a swap partition where applications access remote memory via page faults.  The swap partition is split into a set of 4KB \emph{swap entries}, each mapping to an actual remote memory cell and has a unique entry ID.
Upon a page fault, the kernel uses the swap entry ID contained in the corresponding page table entry (PTE) to locate the swap entry that stores the page.

The first step in handling the fault is to look up the swap cache, which is a set of radix trees, each containing a number of cached and unmapped pages for a block (\eg, 64MB) of swap entries. These pages were either just swapped in due to demand swapping or prefetching, or are about to be swapped out. If a page can be found there, it gets mapped to a virtual page and removed from the swap cache. Otherwise, the kernel needs to perform a \emph{demand swap-in}.

Before issuing the request, the kernel first does \codeIn{cgroup} accounting to understand if there is enough physical memory to swap in the page.
If there is, the kernel issues an RDMA read request, which is then pushed into RDMA's dispatch queue. As the demand swap occurs, the kernel prefetches a number of pages that will likely be needed in the future. This number depends on the swap history at the past few page faults. For example, if the pages fetched follow a sequential or strided pattern, the kernel will use this pattern to fetch a few more pages. If no pattern is found, the kernel reduces the number of prefetched pages until it stops prefetching completely. Once these demand and prefetched pages arrive, they are placed into the swap cache. Their swap entries in remote memory are then freed.

If \codeIn{cgroup} accounting deems that local memory is insufficient for the new page, the kernel uses an LRU algorithm to evict pages. Evicting a page \emph{unmaps} it and pushes it into the swap cache. When memory runs low, the kernel releases existing pages from the swap cache to make room for newly fetched pages. Clean pages can be removed right away and dirty pages must be written back. To write back a page, the swap system must first allocate a swap entry using a free-list-based allocation algorithm. Finally, an RDMA write request is generated and the page is written into the entry via RDMA.

In each remote access, extensive locking is needed for
swap entry allocation\textemdash
shared allocation metadata (\eg, free list) must be protected when multiple applications/threads request swap entries simultaneously.
Although there are active efforts~\cite{patch1, patch2} in the Linux community to optimize swap entry allocation, their performance and scalability is unsatisfactory for cloud workloads (see Appendix \ref{sec:kernelpatches}).

\mysection{Motivating Performance Study \label{sec:motivation}}

To understand the impact of interference, we conducted a study with a set of widely-deployed applications including  Apache Spark~\cite{spark}, Neo4j~\cite{neo4j}, XGBoost~\cite{xgboost} (\ie, a popular ML library), Snappy~\cite{snappy} (\ie, Google's fast compressor/decompressor), as well as Memcached~\cite{memcached}. Spark and Neo4j are managed applications running on the JVM, while the other three are native applications. They cover a spectrum of cloud workloads from data storage through analytics to ML.
In addition, they include both batch jobs (such as Spark) and latency-sensitive jobs (such as Memcached).
Co-running them represents a typical scenario in a modern datacenter where operators fill left-over cores unused by latency-sensitive tasks with batch-processing applications to improve CPU utilization~\cite{datacenter-google}. For example, in a Microsoft Bing cluster, batch jobs are colocated with latency-sensitive services on over 90,000 servers~\cite{perfiso-atc18}. Google also reported that 60\% of machines in their compute cluster co-run at least five jobs~\cite{cpi2-eurosys13}.

We ran these programs, individually vs.~together, on a machine with two Xeon(R) Gold 6252 processors, running Linux 5.5.
Another machine with two Xeon(R) CPU E5-2640 v3 processors and 128GB memory was used for remote memory. Each machine was equipped with a 40 Gbps Mellanox ConnectX-3 InfiniBand adapter and inter-connected by one Mellanox 100 Gbps InfiniBand switch.
Using \codeIn{cgroup}, the same amounts of CPU and local memory resources were given to each application throughout the experiments.
RDMA bandwidth was \emph{not} saturated for both application individual runs and co-runs.
The amount of local memory configured for each application was 25\% of its working set.

\begin{figure}[!ht]
    \centering
      \vspace{-1em}
    \includegraphics[scale=0.33]{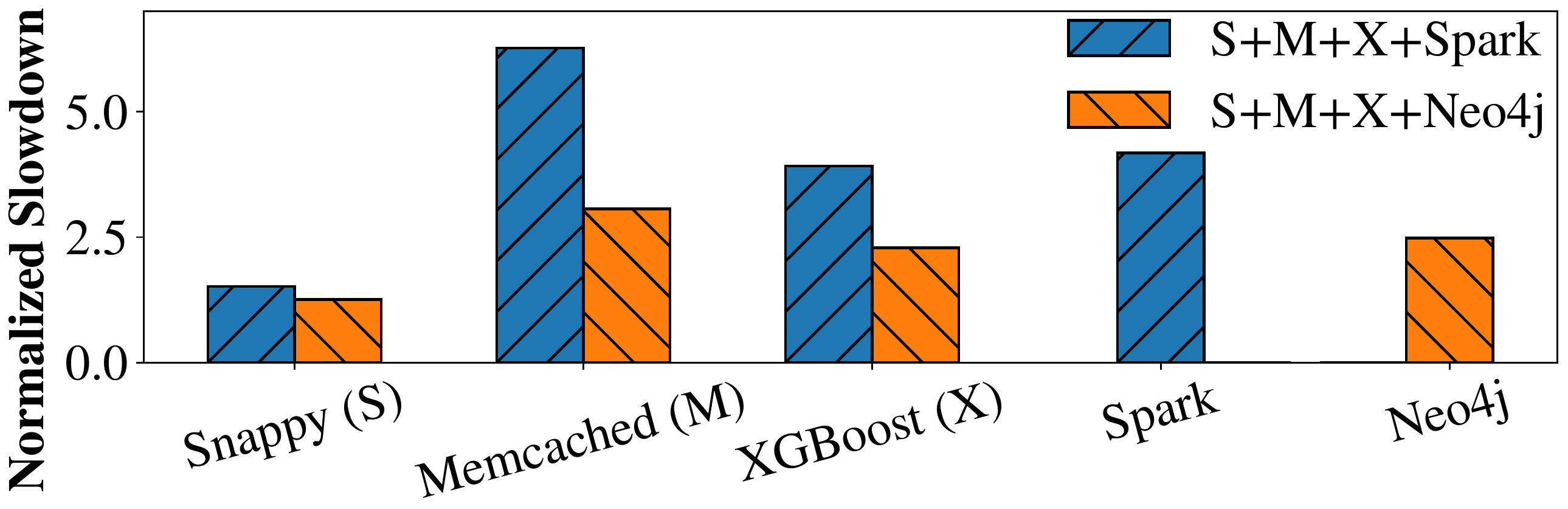}
    \vspace{-1.5em}
    \caption{Slowdowns of co-running applications compared to running each individually.
    \label{fig:multi-app-time}}
    \vspace{-1.em}
\end{figure}

\MyPara{Performance Interference and Degradation.} To understand the overall performance degradation and how it changes with different applications, we used two managed applications: Spark and Neo4j. Figure~\ref{fig:multi-app-time} reports each application's performance degradation when co-running with other applications compared to running alone. The blue/orange bars show the slowdowns when the three native applications co-run with Spark/Neo4j.
Clearly, co-running applications significantly reduces each application's performance.  We observed an overall
\textbf{3.9}/\textbf{2.2$\times$} slowdown when native applications co-run with Spark/Neo4j. Spark persists a large RDD in memory and keeps swapping in/out different parts of the RDD, while Neo4j is a graph database and holds much of its graph data in local memory and thus does not swap as much as Spark.

Another observation is that the impact of interference differs significantly for different applications. Applications that generate high swap throughputs aggressively invade swap and RDMA resources of other applications. In our experiments, Memcached, XGBoost, and Spark all need frequent swaps.
However, Spark runs many more threads ($>$90 application and runtime threads) than Memcached (4) and XGBoost (16), resulting in a much higher swap throughput.
As such, Spark takes disproportionally more resources, leading to severe degradation for Memcached and XGBoost.

\begin{figure}[h!]
  \vspace{-.5em}
    \centering
    \includegraphics[scale=0.33]{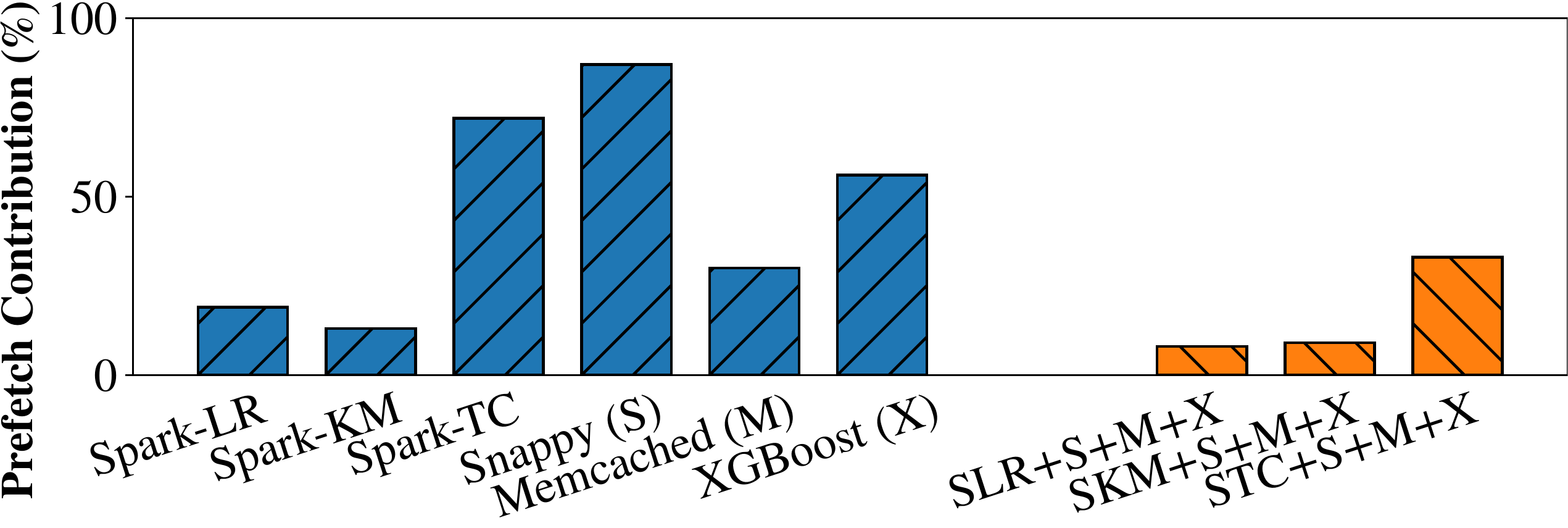}
    \vspace{-1em}
    \caption{Prefetching contribution of Leap: the percentage of page faults served by Leap-prefetched pages (\%).
    \label{fig:multi-app-prefetch}}
    \vspace{-.5em}
\end{figure}

\MyPara{Reduced Prefetching Effectiveness.} Sharing the same prefetching policy reduces the prefetching effectiveness when multiple applications co-run.
Figure~\ref{fig:multi-app-prefetch} reports \emph{prefetching contribution}\textemdash the percentage of page faults served by prefetched pages\textemdash the higher the better; if a prefetched page is never used, prefetching it would only incur overhead. We used Leap~\cite{leap-atc20} as our prefetcher. The left six bars report such percentages for the applications running individually. When applications co-run, the rightmost three bars report the average percentages across applications.  As shown, co-running dramatically reduces the contribution.%

Note that Leap~\cite{leap-atc20} uses a majority-vote algorithm to identify patterns across multiple applications. However, when applications that exhibit drastically different behaviors co-run, Leap cannot adapt its prefetching mechanism and policy to each application. Furthermore, Leap is an aggressive prefetcher\textemdash even if Leap does not find any pattern, it always prefetches a number of contiguous pages.
However, aggressive prefetching for applications such as Spark with garbage collection (GC) is ineffective\textemdash \eg, prefetching for a GC thread has zero benefit and only incurs overhead. Detailed evaluation of prefetching can be found in \S\ref{sec:eval_optimization}.

\begin{figure} [!ht]
    \centering
    \begin{adjustbox}{max width=\linewidth}
    \begin{tabular}{cc}

    \hspace{-1.em}
    \includegraphics[scale=.31]{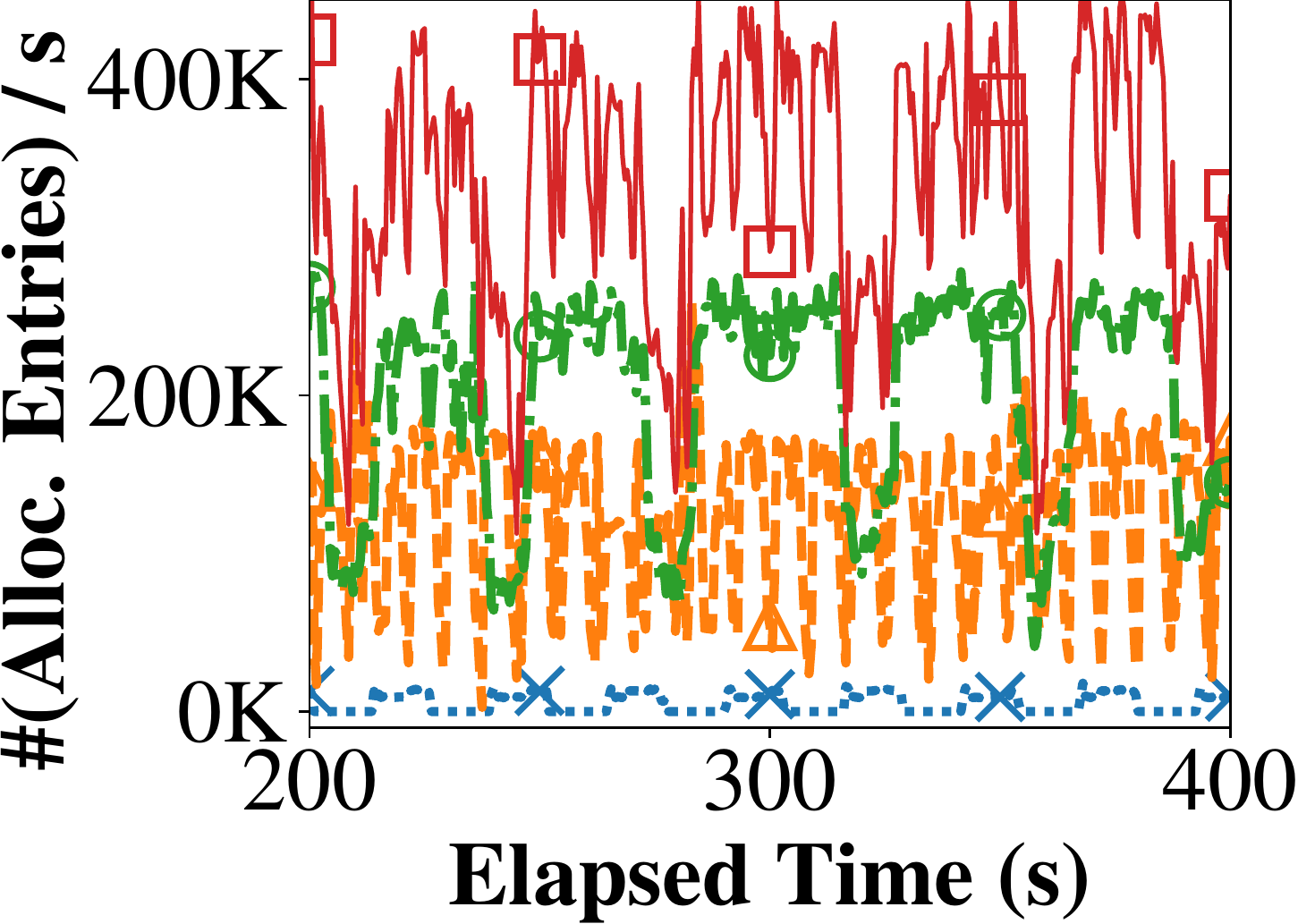} &
    \hspace{-0.7em}
    \includegraphics[scale=.31]{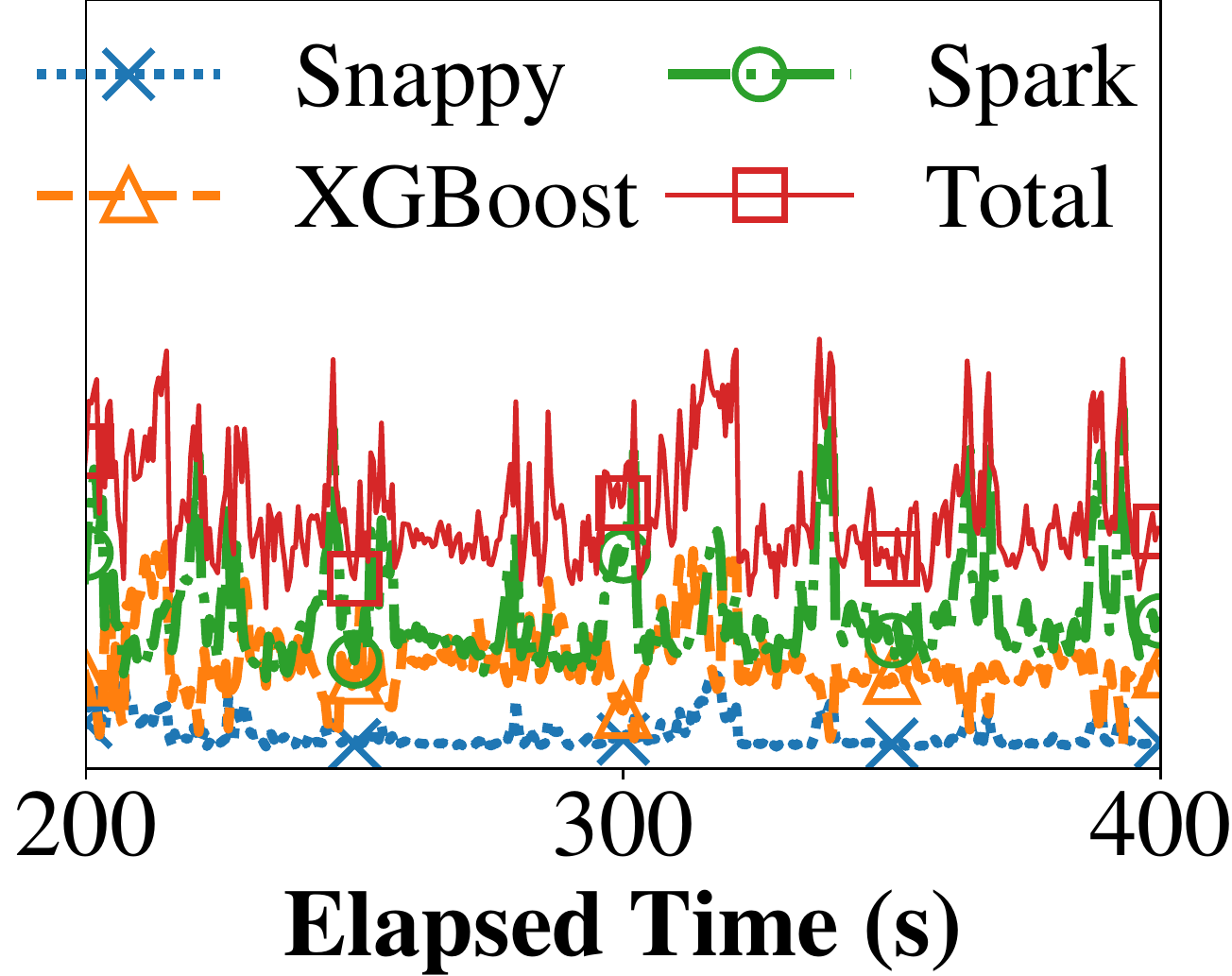}\\[-.2em]
    (a) Running individually. & (b) Co-running.
    \end{tabular}
    \end{adjustbox}
    \vspace{-0.5em}
    \caption{Swap entry allocation throughput when applications run individually (a) and together (b). \label{fig:throughput-degradation}}
    \vspace{-0.5em}
\end{figure}

\begin{figure} [!ht]
    \centering
    \begin{adjustbox}{max width=\linewidth}
    \begin{tabular}{cc}
    \hspace{-1.em}
    \includegraphics[scale=.3]{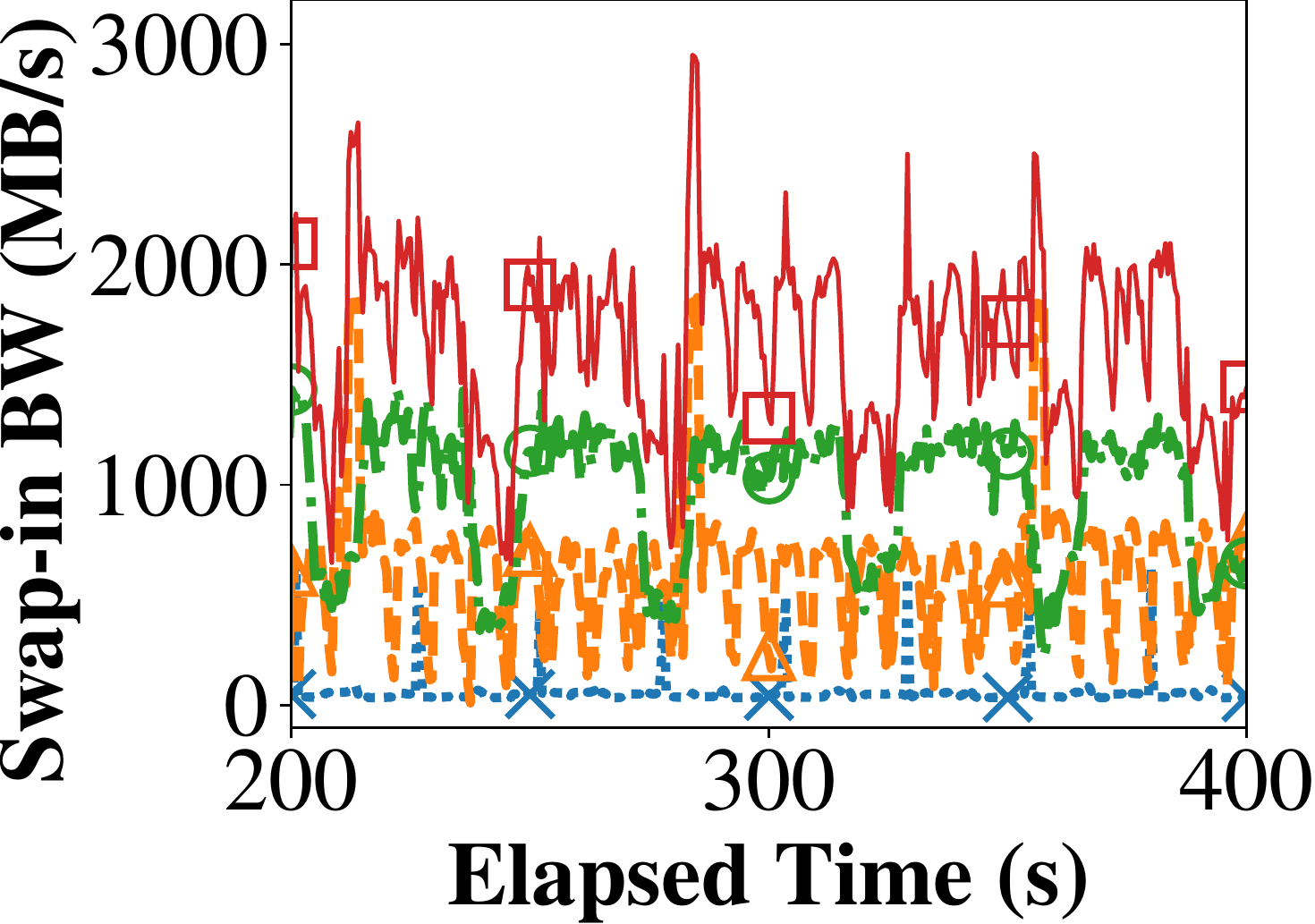} &
    \hspace{-0.7em}
    \includegraphics[scale=.3]{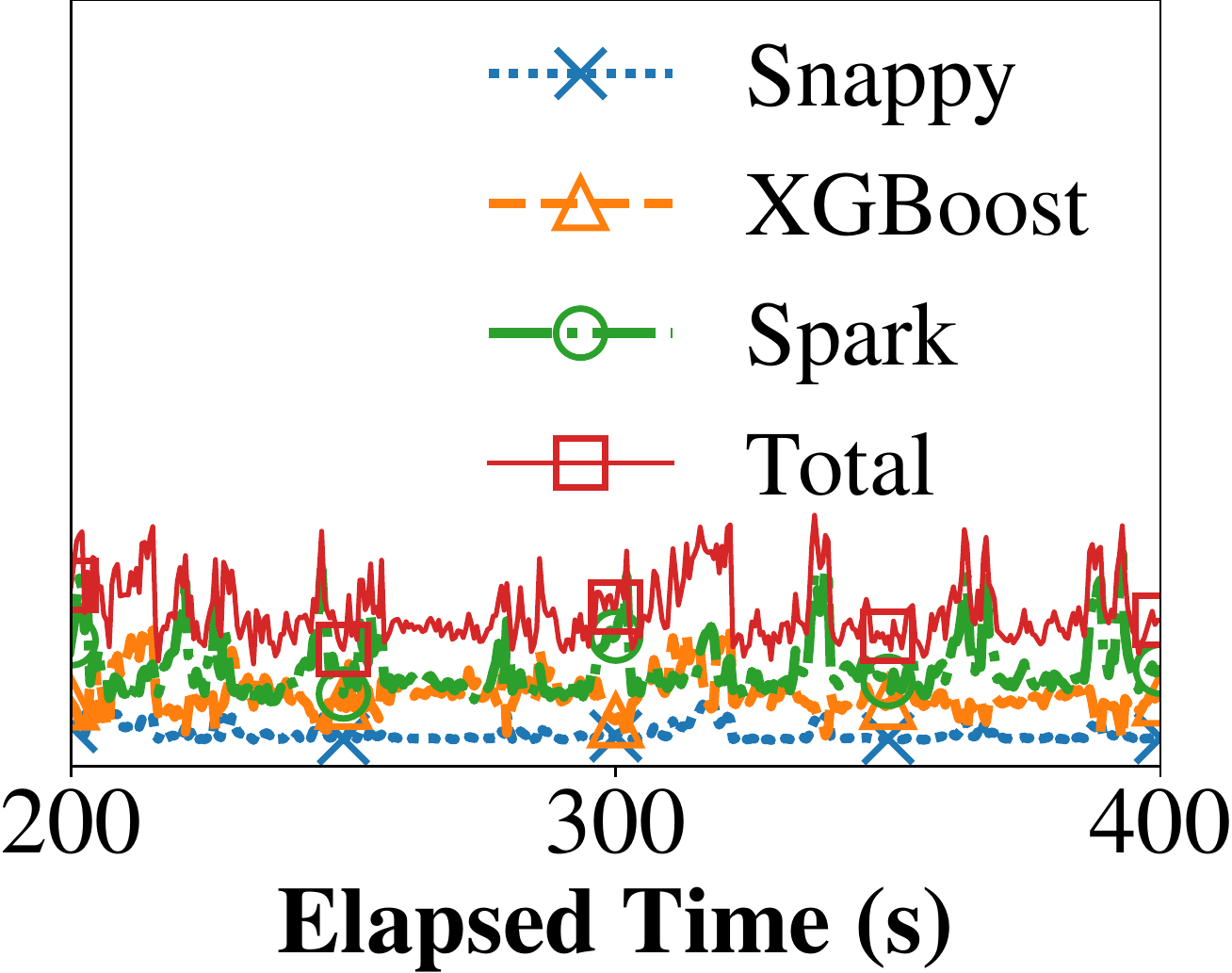}\\[-.2em]
    (a) Running individually. & (b) Co-running.
    \end{tabular}
     \end{adjustbox}
    \vspace{-0.5em}
    \caption{RDMA swap-in bandwidth when applications run individually (a) and together (b).  \label{fig:bandwidth-degradation}}
    \vspace{-.5em}
\end{figure}

\MyPara{Lock Contention.}
We observed severe lock contention in the swap system when applications co-run, particularly at swap entry allocation associated with each swap-out.

We experimented with Spark (Logistic Regression), XGBoost, and Snappy.
Our results show that in windows of frequent remote accesses, co-running applications can spend up to \textbf{70\%} of the window time on obtaining swap entries.
Lock contention leads to significantly reduced swap-entry allocation throughput, reported in Figure~\ref{fig:throughput-degradation}. The \textsf{total} lines in Figure~\ref{fig:throughput-degradation}(a) and (b) show the total throughput (\ie, the sum of each application's allocation throughput). The co-running throughput (b) is drastically reduced compared to the individual run's throughput (a) (\ie, $\thicksim$450Kps to $\thicksim$200Kps).

\MyPara{Reduced RDMA Utilization.}
Figure~\ref{fig:bandwidth-degradation} compares the RDMA read bandwidth (for swap-ins) when applications run individually and together. Similarly, the \textsf{total} line represents the sum of each application's RDMA bandwidth. The total RDMA utilization is constantly below $\thicksim$1000MBps in Figure~\ref{fig:bandwidth-degradation}(b), which is \textbf{3.28$\times$} lower than that in Figure~\ref{fig:bandwidth-degradation}(a) due to various issues (\eg, locking, reduced prefetching, \etc).
The RDMA write bandwidth degrades by an overall of \textbf{2.80$\times$}.

\begin{figure}[h!]
    \centering
    \includegraphics[scale=0.25]{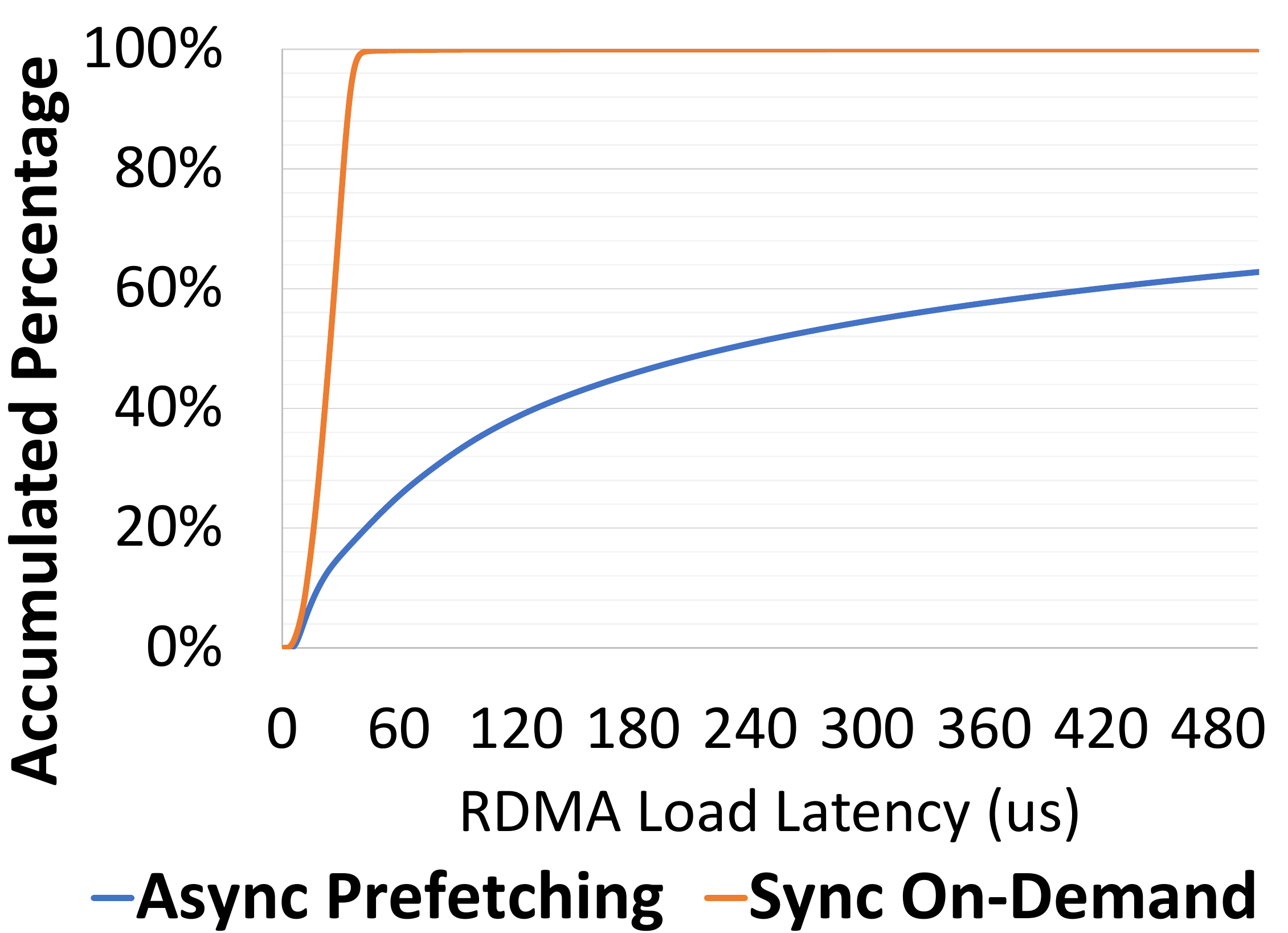}
    \vspace{-0.5em}
    \caption{Latency of prefetching and on-demand swapping.\label{fig:latency_problem_of_async_prefetching}}
    \vspace{-1.em}
\end{figure}

\MyPara{Demand v.s. Prefetching Interference.}
Optimizations such as Fastswap~\cite{fastswap-eurosys20} improve swap performance by dividing the RDMA queue pairs (QP) into sync and async. The high-priority synchronous QP is used for demand swaps, while the low priority async QP is used for prefetching requests. This separation reduces head-of-line blocking incurred by prefetching. However, when applications co-run, this design adds a delay for prefetching. Figure~\ref{fig:latency_problem_of_async_prefetching} depicts the CDF of the latency of RDMA packets from demand and prefetching requests, when the four applications co-run on Leap. As shown, 99\% of the on-demand requests are served within 40$\us$. However, the latency of 36.9\% of prefetching requests is longer than 512$\us$ and it can reach up to 52ms! Long latency renders prefetched pages useless because prefetching is meant to load pages to be used soon. Our profiling shows that among the prefetched pages that are actually accessed by the application, \emph{90\% are accessed within 70$\us$}, indicating that $\thicksim$70\% of the pages prefetched return too late. A late prefetch of a page would subsequently block a demand request of the page when it is accessed by the application. This problem motivates our two-dimensional RDMA scheduling (\S\ref{sec:rdma}).

\begin{table*}
	\centering
	\small
		\begin{tabular}{l|l|l}
			\toprule
			\textbf{Problem Description}  &  \textbf{Performance Impact} & \textbf{\tool's Solution}  \\
			\midrule
		    Unlimited use of swap  &
		    Apps generating higher swap thruput
		    &  Holistic isolation of swap system \\% \cline{1-1} \cline{3-3}
		    and RDMA resources &   use disproportionately more resources & RDMA isolation and scheduling (\S\ref{sec:isolation}, \S\ref{sec:rdma}) \\
		    \midrule
			Lock conten. at swap entry alloc.  & Reduced swap-out thruput & (1) Swap parti. isolation (\S\ref{sec:swap-partition}); (2) adaptive entry alloc. (\S\ref{sec:alloc}) \\
            \midrule
			Single low-level prefetcher & Increased fault-handling latency & Two-tier adaptive prefetching (\S\ref{sec:prefetching}) \\
			\midrule
			prefetching v.s. demand interfere & Increased fault-handling latency & Two-dimensional RDMA scheduling (\S\ref{sec:rdma}) \\
			\bottomrule
		\end{tabular}
	\vspace{-.5em}
		\caption{Summary of major issues and \tool's solution. \label{tab:takeway}}
		\vspace{-2.em}
\end{table*}

\MyPara{Takeaway.} The root cause of performance degradation is that multiple applications, whose resource needs and swap behaviors are widely apart, all run on a global swap system with the same allocator and prefetcher.
Table~\ref{tab:takeway} summarizes these problems, their performance impact, and our solutions.

\mysection{Swap System Isolation~\label{sec:isolation}}

\tool extends \codeIn{cgroup} for users to specify size constraints for swap partition, swap cache, and RDMA bandwidth. We discuss the kernel support to enforce these new constraints, laying a foundation for adaptive optimizations in \S\ref{sec:customization}. %

\MyPara{Swap Partition Isolation. \label{sec:swap-partition}}
In Linux, remote memory is managed via a swap partition interface, shared by all applications. If there are multiple available swap partitions, they are used in a \emph{sequential manner} according to their priorities. As a result, data of different applications are mixed and stored in arbitrary locations.

\tool separates remote memory of each \codeIn{cgroup} to isolate capacity and performance.  
The user creates a \codeIn{cgroup} to set a size limit of remote memory for an application. \tool allocates remote memory in a demand-driven manner\textemdash upon a pressure in local memory, \tool allocates remote memory and registers it as a RDMA buffer. \tool enables per-\codeIn{cgroup} swap partitions by creating a swap partition interface and attaching it to each \codeIn{cgroup}. For each \codeIn{cgroup}, a separate swap-entry manager is used for allocating and freeing swap entries. Swap entry allocation can now be charged to the \codeIn{cgroup}, which controls how much remote memory each application can use. Our adaptive swap entry allocation algorithm is discussed in \S\ref{sec:alloc}.

\tool explicitly enables a private swap cache for each \codeIn{cgroup} (a default value of 32MB), whose size is charged to the \emph{memory budget} specified in the \codeIn{cgroup}. As a result, the size of an application's swap cache changes in response to its own memory usage, without affecting other applications.

For each demand swap-in, \tool first checks the \codeIn{mapcount} of the page, which indicates how many processes this page has been mapped to before. If the page belongs only to one process,
 it is placed in its private swap cache. Otherwise, it has to be placed in a global swap cache (discussed shortly).
To release pages (\eg, when the application's working set increases, pushing the boundary of the swap cache), \tool scans the swap cache's page list,
releasing a batch of pages to shrink the cache. 

\MyPara{RDMA Bandwidth Isolation.\label{sec:rdma_isolation}}
For each \codeIn{cgroup}, \tool isolates RDMA bandwidth with a set of \emph{virtual} RDMA queue pairs (VQPs) and a centralized packet scheduler. Users can set the swap-in/swap-out RDMA bandwidth of a \codeIn{cgroup} with our extended interface.
Our RDMA scheduler works in two dimensions. The \emph{first dimension} schedules packets across applications, while the \emph{second dimension} schedules on a per-application basis\textemdash each \codeIn{cgroup} has its \emph{sub-scheduler} that schedules packets that belong to the \codeIn{cgroup} between demand swapping and prefetching.  

VQPs are high-level interfaces, implemented with lock-free linked lists. Each \codeIn{cgroup} pushes its requests to the head of its VQP, while the scheduler pops requests from their tails. At the low level, our scheduler maintains three physical queue pairs (PQP) per core, for \emph{demand swap-in}, \emph{prefetching}, and \emph{swap-out}, respectively.  The scheduler polls all VQPs and forwards packets to the corresponding PQPs, using a \emph{two-dimensional} scheduling algorithm (see \S\ref{sec:rdma}).

\MyPara{Handling of Shared Pages.\label{sec:shared}}
Processes can share pages due to shared libraries or memory regions. These pages cannot go to any private swap cache. \tool maintains a global swap partition and cache for shared pages. When a page is evicted and ummapped, \tool checks its  \codeIn{mapcount} and adds it to the global swap cache if  %
the page is shared between different processes.
All pages in the global swap cache will be eventually swapped out to the global partition using the original lock-based allocation algorithm. Conversely, pages swapped in (and prefetched) from the global swap partition are all placed into the global swap cache. 
For typical cloud applications such as Spark, Cassandra and Neo4j, the number of shared pages is much smaller than process-private pages, using locks in a normal way would not incur a large overhead. 
We cannot charge applications' \codeIn{cgroups} for pages in the global swap cache, because which process(es) share these pages is unknown before they get mapped into processes' address spaces. \tool allows users to create a special \codeIn{cgroup}, named \codeIn{cgroup-shared}, to limit the size of the global swap cache/partition. %

One limitation of our \codeIn{cgroup}-based approach is that \codeIn{cgroup} can only partition resources statically while applications' resource usage may change from time to time and static partitioning could lead to resource underutilization. However, the focus of this paper is to ensure isolation and future work could incorporate max-min fair allocation to improve resource utilization. 
\mysection{Isolation-Enabled Swap Optimizations\label{sec:customization}}
On top of the isolated swap system, we develop three optimizations, which dynamically adapt their strategies to each application's resource patterns and semantics.

\subsection{Adaptive Swap Entry Allocation \label{sec:alloc}}

As discussed in \S\ref{sec:motivation}, swap entry allocation suffers from severe lock contention under frequent remote accesses\textemdash allocation is needed at every swap-out.  To further motivate, we use a simple experiment by running Memcached alone on remote memory with different core numbers. As the number of cores increases, the average entry allocation time grows super-linearly\textemdash it grows from 10\us under 16 cores quickly to 130\us under 48 cores due to increased lock contention (see Figure~\ref{fig:5.14-memcached}). Creating a per-application swap partition mitigates the problem to a certain degree. However, applications like Spark run more than 90 threads; frequent swaps in these threads can still incur significant locking overhead.

To further reduce contention, we develop a novel swap entry allocator that adapts allocation strategies in response to each application's own memory access/usage. Our first idea is to enable a \emph{one-to-one} mapping between pages and swap entries. At the first time a page is swapped out, we allocate a new swap entry using the original (lock-protected) algorithm. Once the entry is allocated, \tool writes the entry ID into the page metadata (\ie, \codeIn{struct page}). This ID remains on the page throughout its life span. As a result, subsequent swap-outs of the page can write data directly into the entry corresponding to this ID. We pay the locking overhead \emph{only once} for each page at its first swap-out.

This approach requires a swap entry to be reserved for each page. For example, if the local memory size is S and the remote memory allocation is 3S, with one-to-one mapping the remote memory allocation would be 4S (\ie, each page residing in local memory also has a remote page, resulting in a 33\% overhead). However, this overhead may not be necessary. 
For example, modern applications exhibit strong epochal behaviors. Under the original allocator, swap entries for pages accessed in one epoch can be reused for those in another epoch. Under this approach, however, all pages in all epochs must have their dedicated swap entries throughout the execution, which can lead to an order-of-magnitude increase in remote memory usage. 

Our key insight is: 
we should trade off \emph{space for time} if an application has much available swap space, but \emph{time for space} when its space limit is about to be reached.  
As such, when the remote memory usage is about to reach the limit specified in \codeIn{cgroup} (\ie, 75\% in our experiments), \tool starts removing reservations to save space. The next question is which pages we should consider first as our candidates for reservation removal. Our idea is that we should first consider ``hot pages'' that always stay in local memory and are rarely swapped. This is because hot pages (\ie, data on such pages are frequently accessed) are likely to stay in local memory for a long time; hence, locking overhead is less relevant for them. On the contrary, ``cold'' pages whose accesses are \emph{spotty} 
are more likely to be swapped in/out and hence swap efficiency is critical. Here ``hot'' and ``cold'' pages are relatively defined as they are specific to execution stages\textemdash a cold page swapped out in a previous stage can be swapped in and become hot in a new stage. 

To this end, we develop an \emph{adaptive allocator}. \tool starts an execution by reserving swap entries for \emph{all} pages to minimize lock contention. Reservation removal begins when remote-memory pressure is detected.  \tool adaptively removes reservations for hot pages. %
We detect hot pages \emph{for each application} by periodically scanning the application's \emph{LRU active list}\textemdash pages recently accessed are close to the head of the active list. Each scan identifies a set of pages from the head of the list; a page is considered ``hot'' if it appears in a consecutive number of sets recently identified. 

Removing the reservation for a hot page can be done by (1) removing the entry ID from the page metadata and (2) freeing its reserved swap entry in remote memory, adding the entry back to the free list. Once a hot page becomes cold and gets evicted, it does not have a reservation any more, and hence, it goes through the original (lock-protected) allocation algorithm to obtain an entry. In this case, the page receives a new swap entry and remembers this new ID in its metadata.

\begin{figure}[h!]
    \centering
   \vspace{-.8em}
    \includegraphics[scale=.35]{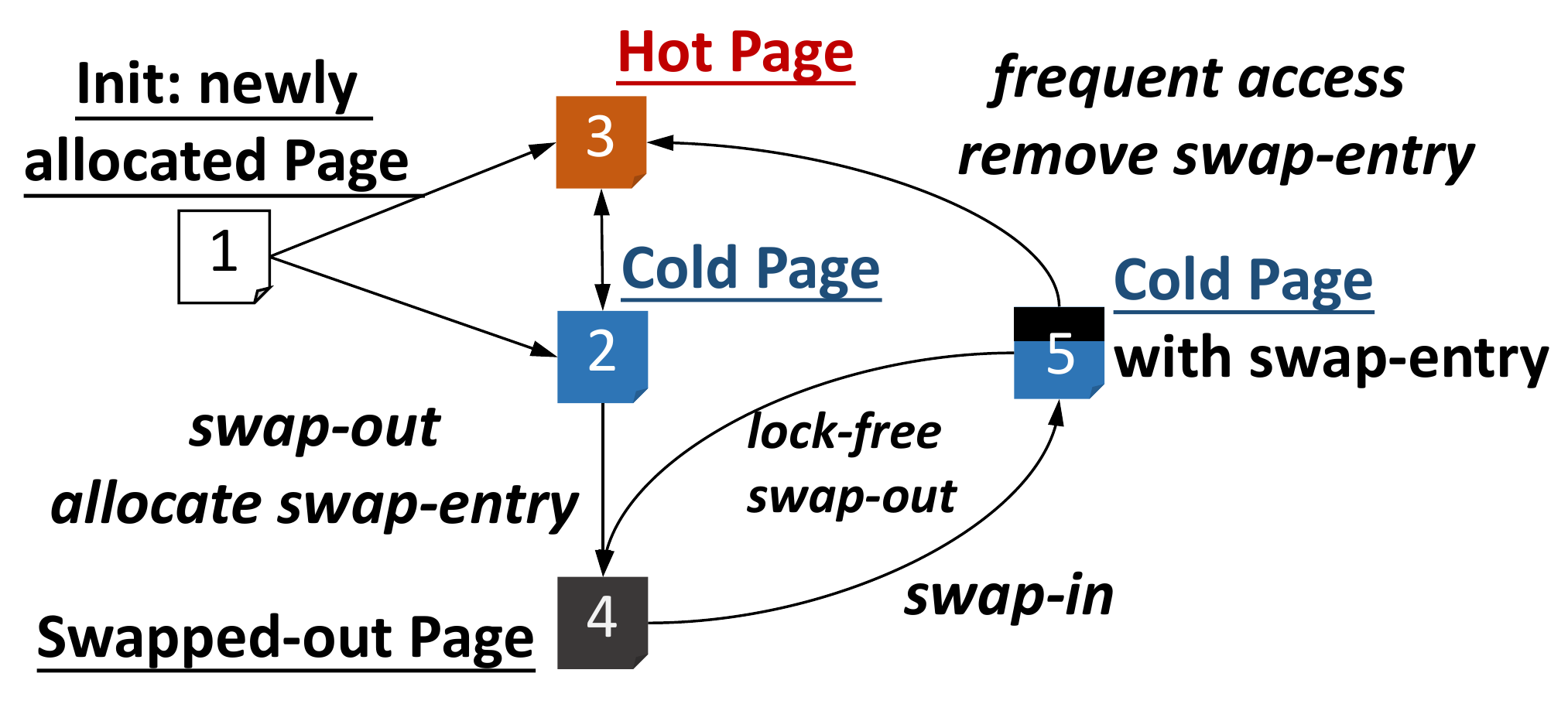}
    \vspace{-0.5em}
    \caption{FSM describing our page management when remote-memory pressure is detected.}
    \label{fig:state-machine}
\vspace{-1em}    
    \end{figure}

Figure~\ref{fig:state-machine} shows the page state machine, which describes the page handling logic. A cold page (to be evicted) can be in one of the two states: state 2 and state 5. A page comes to state 2 if it is (1) a brand new page that has never been swapped out or (2) previously a hot page but has not been accessed for long. Once it reaches state 2, the page does not have a reserved swap entry ID and hence, swapping out this page goes through the normal allocation path. In the case of swap-in (state 5), the swap entry ID is already remembered on the page. The next swap-out will directly use this entry and be lock-free. If the page becomes hot (from state 5 to 3), \tool removes the entry ID and releases the entry reservation. The entry is then added back to the free list.  

\MyPara{Performance Analysis.}
To understand the performance of the adaptive entry allocation algorithm, let us consider the following two scenarios. In the first scenario, the application performs uniformly random accesses. 
As a result, \tool cannot clearly distinguish hot/cold pages, and thus randomly cancels their reservations.
However, due to the random process, when a page is swapped out, it has a certain probability of still possessing a reserved swap entry (depending on the ratio of remaining reservations) and hence \tool{} can still improve the allocation performance.

In the second scenario, the application follows a repetitive pattern of accessing a page a few times (making it hot) and then moving on to accessing another page; it will not come back to the page in a long while. Under our allocation algorithm, every page will be identified as a hot page, leading to the cancellation of its reservation. However, each page will be swapped out when it is cold enough; at each swap-out, the page has to go through the original allocation algorithm. 
This is the worst-case scenario, and even in this case, \tool{} has the same (worst-case) performance as the original Linux allocator, which allocates an entry at each swap-out. 

Some of the recent patches submitted to the Linux community also attempt to reduce lock contention for swap entry allocation. A detailed description of how \tool differs from these patches can be found in Appendix \ref{sec:kernelpatches}.

\subsection{Two-tier Adaptive Prefetching\label{sec:prefetching}}
\vspace{-.5em}
\MyPara{Problems with Current Prefetchers.} Current prefetchers all focus on low-level (streaming or strided) access patterns. While such patterns exist widely in native array-based programs, applications written in high-level languages such as Python and Java are dominated by reference-based data structures\textemdash operations over such data structures involve large amounts of pointer chasing, making it hard for current prefetchers to identify clear patterns. 

Furthermore, cloud applications such as Spark are heavily multi-threaded. Modern language runtimes, such as the JVM, run an additional set of auxiliary threads, \eg, for GC or JIT compilation. How these user-level threads map to kernel threads is often implemented differently in different runtimes. Consequently, kernel prefetchers such as Leap~\cite{leap-atc20} cannot distinguish patterns from different threads.%

\begin{figure}[t]
    \centering
    \includegraphics[scale=.43]{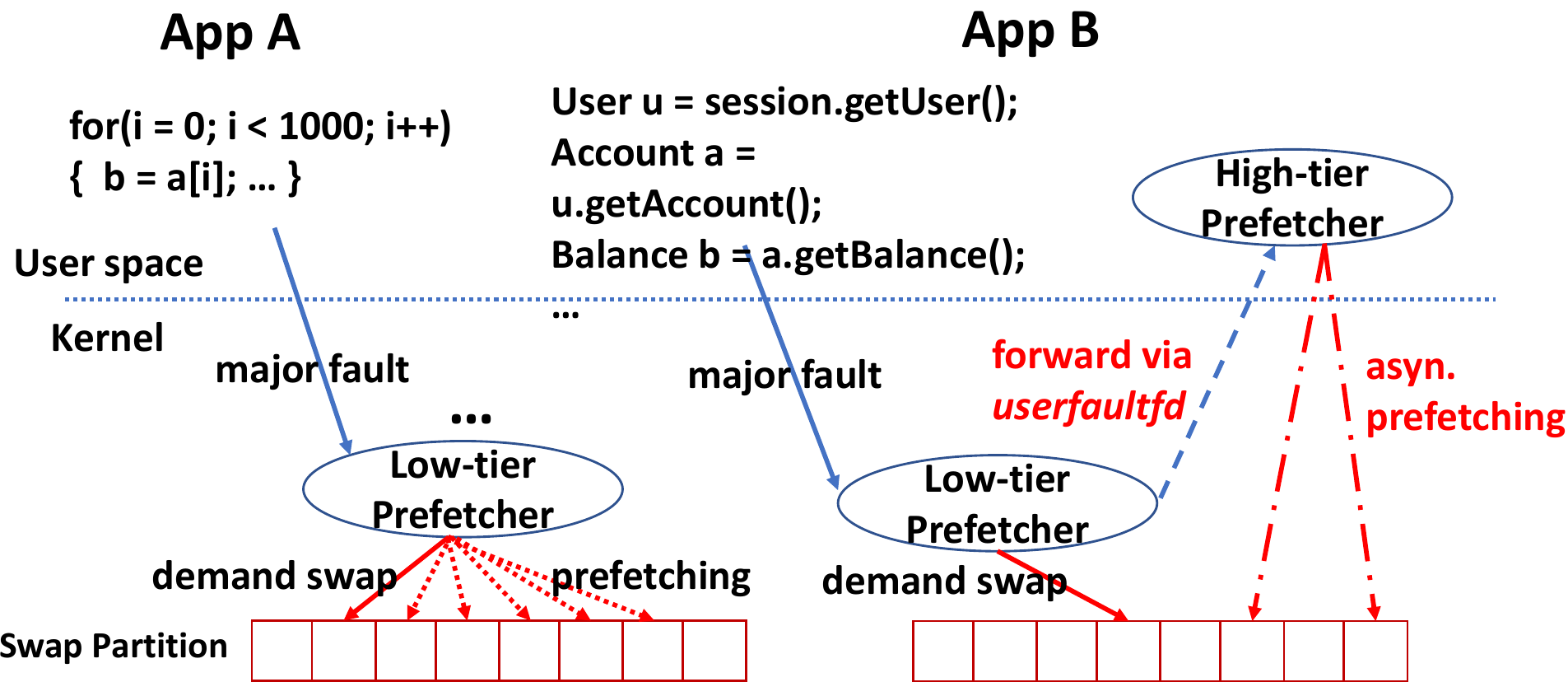}
    \vspace{-0.5em}
    \caption{\tool's two-tier prefetcher: App A is an array-based program while B is a modern web application that uses reference-based data structures. The low-tier prefetcher successfully prefetches pages for A, but not for B. Hence, \tool forwards the addresses up to B's high-tier prefetcher. \label{fig:prefetcher}}
    \vspace{-1.5em}
\end{figure}

To develop an adaptive prefetcher, \tool employs a two-tier design, illustrated in Figure~\ref{fig:prefetcher}. At the low (kernel) tier, \tool uses an existing kernel prefetcher that prefetches data for each application into its own private swap cache (unless data comes from the global swap partition). A kernel prefetcher is extremely efficient and can already cover a range of (array-based) applications. For applications whose accesses are too complex for the kernel prefetcher to handle, we forward the addresses up to the application level, letting the application/runtime analyze semantic access patterns at the level of threads, references, arrays, \etc

\MyPara{Prefetching Logic.}  In \tool, we adopt the sync/async separation design in Fastswap~\cite{fastswap-eurosys20}, which prevents head-of-line blocking. As stated earlier, we use three PQPs per core, one for swap-out, one for (sync) demand swap-in, and one for (async) prefetching. 
\tool polls for completions of critical (demand) operations, while configuring \emph{interrupt completions}
for asynchronous prefetches. %

\tool determines whether to use an application-tier prefetcher based on \emph{how successful kernel-tier prefetching is}. If the number of pages prefetched for an application is lower than a threshold at the most recent N (=3 in our evaluation) faults consecutively, \tool starts forwarding the faulting addresses up to the application-tier prefetcher (discussed shortly) although the kernel-tier prefetcher is still used as the first-line prefetcher.

\tool stops forwarding whenever the kernel-tier prefetcher becomes effective again. Our key insight is: the kernel-tier prefetcher is efficient  without needing additional compute resources (as it uses the same core as the faulting thread), while the application-tier prefetcher needs extra compute resources to run. As such, we disable application-tier prefetchers as long as the kernel-tier prefetcher is effective. To pass a faulting address to the application, we modify the kernel's \codeIn{userfaultfd} interface, allowing applications to handle faults at the user space. Our modification makes the kernel forward the faulting address only if the kernel's prefetcher continuously fails to prefetch pages.

\MyPara{Runtime Support for Application-tier Prefetching.} A major challenge is how to develop application-tier prefetchers. On the one hand, application-tier prefetchers should conduct prefetching based on \emph{application semantics}, of which the kernel is unaware. On the other hand, 
application developers may not be familiar with a low-level activity like prefetching;
understanding memory access patterns and developing prefetchers can be a daunting task for them. 

Our insight is: applications that benefit from application-tier prefetching are mostly written in high-level languages and run on a managed runtime such as the JVM.  
Inspired by previous work on using language runtime to solve memory efficiency problems for data analytics applications~\cite{yak-osdi16,gerenuk-sosp19,facade-asplos15,skyway-asplos18, taurus-asplos16}, \tool currently supports application-tier prefetching for the JVM as a platform. However its support could be easily extended to other managed runtimes for high-level languages like Go and C\#. Leveraging language runtime solves both problems discussed above\textemdash it has access to semantic information such as how objects are connected and the number of application threads; furthermore, the burden of developing an application-tier prefetcher is shifted from application developers to runtime developers. Thus, it is not necessary to supply a custom application-tier prefetcher per application, but define it once for each language runtime.

In this work, we develop an application-tier prefetcher in Oracle's OpenJDK as a proof-of-concept. It works for all (Java, Scala, Python, \etc) programs that run on the JVM. Our JVM-based prefetcher considers two \emph{semantic patterns}: (1) \emph{reference-based} (\ie, accessing an object brings in pages containing objects referenced by this object) and (2) \emph{thread-based} (\ie, accesses from different application threads are separately analyzed to find patterns).

For (1), we modify the JVM to add support that can quickly find, from a faulting address, \emph{the object} in which the address falls. We use write barrier, a piece of code instrumented by the JVM at each object field write, as well as the garbage collector to record references between pages. For example, for each write of the form \codeIn{a.f=b}, if the objects referenced by $a$ and $b$ are on different page groups, we record an edge on a \emph{summary graph} where each node represents a consecutive group of pages and each edge represents references between groups. During prefetching, we traverse the graph from the node that represents the accessed page and prefetch pages that can be reached within 3 hops. The traversal does not follow cycles and its overhead is negligible. This approach is suitable for applications that store a large amount of data in memory, such as Spark and Cassandra.

For (2), we leverage the JVM's user-kernel thread map. For each faulting address, \tool additionally forwards the thread information (\ie, pid) to the JVM, which consults  the map to filter out non-application (\eg, GC, compilation, \etc) threads and segregate addresses based on Java threads (as opposed to kernel threads). Segregated addresses allow us to  analyze (sequential/strided) patterns on a per-thread basis (using Leap's majority-vote algorithm~\cite{leap-atc20}). Once patterns are found, the prefetcher sends the prefetching requests to the kernel via \codeIn{async\_prefetch}.

For native programs that directly use kernel threads (\eg, \codeIn{pthread}), the thread information is straightforward and immediately visible to \tool. We can easily segregate addresses accessed from different threads and analyze patterns based upon addresses from each individual thread.

\MyPara{Policy.} To improve effectiveness, the JVM uses a search tree to record information about large arrays.
Upon the allocation of an array whose size exceeds a threshold (\ie, 1MB in our experiments), the JVM records its starting address and size into the tree. The JVM runs a daemon prefetching thread. Once receiving a sequence of faulting addresses, we determine which semantic pattern to use based on \emph{how many application threads are running} and \emph{whether the faulting addresses fall into a large array}. If there are many threads and the faulting addresses fall into arrays, the JVM uses (2) to find per-thread patterns. If either condition does not hold, the JVM uses (1) to prefetch based on references. 
For native applications, we only enable (2), as we observed that our native programs do not use many deep data structures.%

\vspace{-.2em}
\mysubsection{Two-Dimensional RDMA Scheduling\label{sec:rdma}}

To provide predictable performance for applications sharing RDMA resources, our RDMA scheduling algorithm should provide four properties: (1) weighted fair bandwidth sharing~\cite{max-min-fairness, wfq-demers-sigcomm89} across applications; (2) high overall utilization; (3) treating demand and prefetching requests with different priorities; and (4) timely handling of prefetching requests. 

\tool performs two-dimensional scheduling by extending existing techniques. \tool uses max-min fair scheduling to assign bandwidth across applications, and priority-based scheduling with \emph{timeliness} to schedule prefetching and demand requests within each application. Although these scheduling techniques are not new themselves, \tool combines them in a unique way to solve the interference problem.  \tool maintains three PQPs on each core, respectively, for swap-outs, demand swap-ins, and prefetching swap-ins. Swap-outs are only subject to fair scheduling while swap-ins are subject to both fair and priority-based scheduling.

\MyPara{Vertical: Fair Scheduling.} Under max-min fairness, each application receives a fair share of bandwidth. If there is extra bandwidth, we give it to the applications in the reverse order of their bandwidth demand until bandwidth is saturated. 
The high overall utilization of bandwidth is achieved by redistributing unconsumed bandwidth proportionally to the weights of unsatisfied applications.
\tool implements weighted fair queuing with virtual clock~\cite{pgps-parekh-1993, wfq-demers-sigcomm89, virtualclock-zhang1989new}.

\MyPara{Horizontal: Priority Scheduling with Timeliness.} %
Within each \codeIn{cgroup}, \tool schedules demand requests with a higher priority than prefetching requests. However, this could lead to long latency for prefetching requests. 
To bound the latency of prefetching, our scheduler employs a history-based heuristic algorithm to identify and drop outdated prefetching requests. In particular, \tool maintains the \emph{timeliness distribution} of prefetched pages per \codeIn{cgroup}. 
Timeliness is a metric that measures the time between a page being prefetched and accessed. We attach a timestamp to each request when pushing it into a VQP. The scheduler maintains packets statistics on-the-fly to estimate the round-trip latency and arrival time of each prefetching request. Requests are dropped if the estimated arrival time exceeds the estimated timeliness threshold. 

Special care must be taken to drop prefetching requests. Before issuing a prefetching request, the kernel creates a page in the swap cache and sets up its corresponding PTE. The page is left in a \emph{locked} state until its data comes back.  However, a thread that accesses an address falling into the page may find this locked page in the swap cache and block on it. Dropping prefetching requests may cause the thread to hang. To solve the problem, we detect threads that block on prefetching requests for too long and generate new \emph{demand requests} for them.

We rely on a per-entry timestamp to efficiently detect threads that block on prefetching requests. In \tool{}, we attach a timestamp field to the swap entry metadata. \tool{}'s scheduler records the timestamp every time it enqueues a prefetching request into VQP. If another thread faults on the same page later, it will retrieve the same swap entry from the PTE. If the swap entry contains a timestamp, the faulting thread knows that a prefetching request has already been issued. 
Next, the faulting thread calculates the time elapsed since the timestamp, and compares it with a timeout threshold (maintained by the RDMA scheduler based on page-fetching latencies). If it exceeds the timeout threshold, the faulting thread drops the prefetching request. The drop operation is elaborated below:

Before issuing each (demand or prefetching) request, the kernel first allocates a physical page in the swap cache and locks the page until the request returns. Upon the return of the data, the data is written into the page; the page is  unlocked and mapped into the page table. In order to safely drop a request, we add another field \emph{valid} in the swap entry metadata, indicating whether the prefetching request on the go is valid. Once a faulting thread identifies a delayed prefetching request (by using the timestamp as discussed above), it sets the \emph{valid} field in the swap entry to \codeIn{false} and then creates a new physical page in the swap cache. The thread goes ahead and issues another (demand) I/O request based on this new page. When the delayed prefetching request returns, it checks the \emph{valid} field and discards itself once it sees the false value. The field is then set back to \codeIn{true}. 

When a demand request is issued, \tool clears the timestamp field in its corresponding swap entry. If a thread faults on the same page, it will block on the request instead of issuing a new one due to the empty timestamp (indicating that the request on the go is a demand one).

\mysection{Evaluation\label{sec:eval}}

It took us 17 months to implement \tool in Linux 5.5. The application-tier prefetcher was implemented in OpenJDK 12.

\begin{table}[h]
\vspace{-0.5em}
\centering
\begin{adjustbox}{max width=\linewidth}

\begin{tabular}{lllr}
    \toprule
    \textbf{Application} & \textbf{Workload} & \textbf{Dataset} & \textbf{Size / ($|E|$, $|V|$)}\\
    \midrule
    \rowcolor[HTML]{EFEFEF}
    \textbf{Managed} & & &\\
    Cassandra & 5M read, 5M insert & YCSB\cite{cooper-ycsb-ds}   & 10M records\\
    Neo4j & PageRank & Baidu\cite{wikipedia-ds} & (17M, 2M) \\
    Spark
     & PageRank (\textbf{SPR}) & Wikipedia\cite{wikipedia-ds} & (57M, 1.5M)\\
     & KMeans (\textbf{SKM}) & Wikipedia\cite{wikipedia-ds} & 188M points\\
     & Logistic Regression (\textbf{SLR})  & Wikipedia\cite{wikipedia-ds} & 188M points\\
     & Skewed Groupby (\textbf{SSG}) &  synthetic & 256K records\\
     & Triangle Counting (\textbf{GTC}) & synthetic & (1.5M, 384K) \\
    MLlib & Bayes Classifiers (\textbf{MBC}) & KDD~\cite{kdddata} & 1.5M instances\\
    GraphX & Connected Components (\textbf{GCC)} & Wikipedia\cite{wikipedia-ds}  & (188M, 9M)\\
    & PageRank (\textbf{GPR}) & Wikipedia\cite{wikipedia-ds}  & (188M, 9M)\\
    & Single Src. Shortest Path (\textbf{GSP})  & synthetic & 2M vertices\\
    \toprule
    \rowcolor[HTML]{EFEFEF}
    \textbf{Native} &  & & \\
    XGBoost & Binary Classification & HIGGS\cite{higgs-ds}  & 22M instances\\
    Snappy & Compression &  enwik9~\cite{compress-enwik9-ds}  & 16GB \\
    Memcached & 45M gets, 5M sets & YCSB\cite{cooper-ycsb-ds} & 10M records\\
    \bottomrule
\end{tabular}
\end{adjustbox}
\vspace{-0.5em}
\caption{\label{tab:spark_programs} Programs and their workloads.}
\vspace{-0.5em}
\end{table}

\MyPara{Setup.} We included a variety of cloud applications in our experiments, including managed (Java) applications such as Spark~\cite{spark}, Cassandra~\cite{cassandra} (a NoSQL database), Neo4j~\cite{neo4j} (a graph database), as well as three native applications: XGBoost~\cite{xgboost}, Snappy~\cite{snappy}, and Memcached~\cite{memcached}. %
Spark, Cassandra, Neo4j, Memcached, and XGBoost are multi-threaded while Snappy is single-threaded. The Spark applications span popular libraries such as GraphX and MLlib.

We co-ran different combinations of programs. The same application in different combinations receives the same amount of local (CPU and memory) resources.
To simplify performance analysis, we let each combination of applications co-run contain one managed (Spark, Cassandra, or Neo4j) application and the three native programs, which consume less resources.
These experiments were conducted on two machines, one used to execute applications and a second to provide remote memory. The configurations of these machines was reported earlier in \S\ref{sec:motivation}.
We carefully configured Linux with the following configuration to achieve the best performance for Linux: (1) SSD-like swap model, (2) per-VMA prefetching policy, and (3) cluster-based swap entry allocation.
We disabled hyper-threads, CPU C-states, dynamic CPU frequency scaling, transparent huge pages, and the kernel's mitigation for speculation attacks.

For each combination, we limited the amounts of CPU resources for the managed application, XGBoost, Memcached, and Snappy to be 24, 16, 4, and 1 core(s).
For local memory, we used two ratios: 50\% and 25\%, meaning each application has 50/25\% of its working set locally.
When using \tool, we additionally limited the sizes of swap partitions in such a way that for each application the total size of its swap partition and assigned local memory is slightly larger than its working set. In doing so, each application has just enough (local and remote) memory to run and reservation cancellation (\S\ref{sec:alloc}) is triggered in all executions.

The swap cache size for each application starts at 32MB and changes dynamically. The global swap cache size (configured by \codeIn{cgroup-share}) was also set to 32MB.
Canvas uses max-min fair scheduling to assign bandwidth across applications, and their initial weights are proportional to their swap partition assignments.
We ran each application 10 times. Their average execution times (with error bars) are reported in all experiments throughput this section.

\begin{figure} [h!]
    \centering
    \vspace{-0.2em}
        \includegraphics[scale=.39]{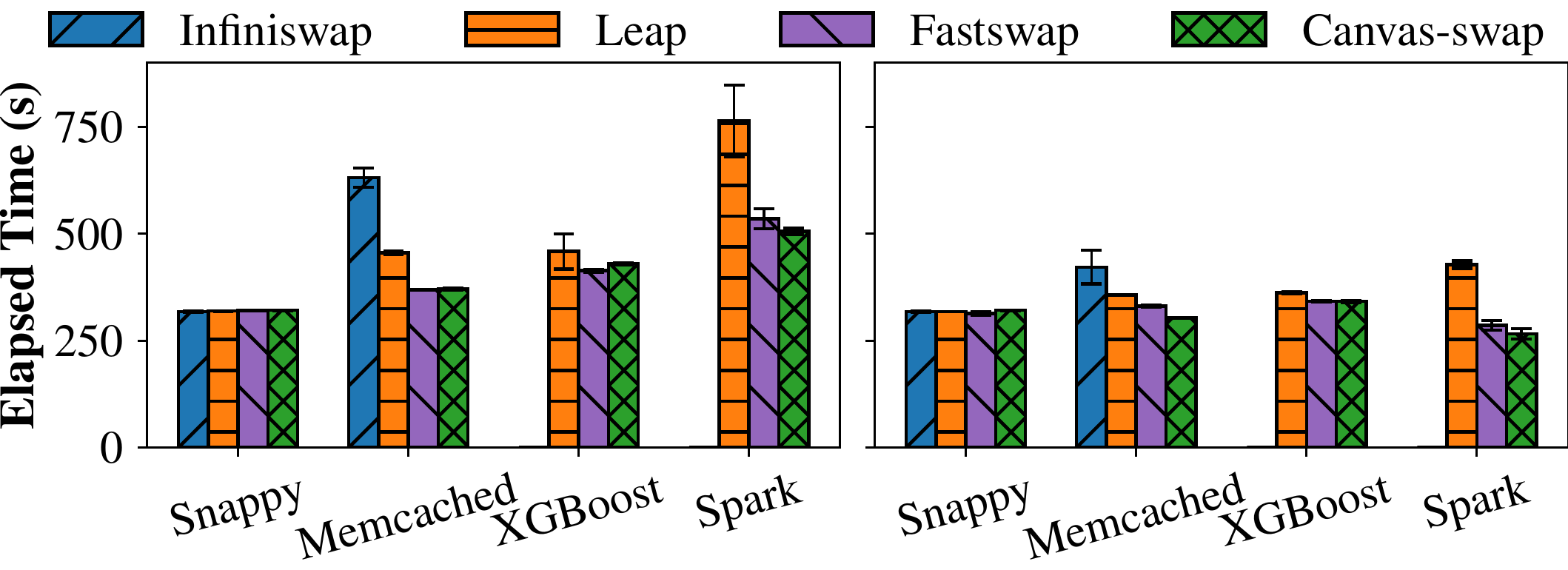}
    \begin{tabular}{cc}
        \hspace{1.2em} (a) 25\% local memory. \label{fig:25-baseline-swap} &
        \hspace{1.5em} (b) 50\% local memory. \label{fig:50-baseline-swap}
    \end{tabular}
    \vspace{-0.4em}
    \caption{Performance of different swap systems. \label{fig:baseline-swap-systems}}
    \vspace{-1.2em}
\end{figure}

\begin{figure*} [t]
    \centering
      \begin{tabular}{c}
        \hspace{-0.5em} \includegraphics[scale=.4]{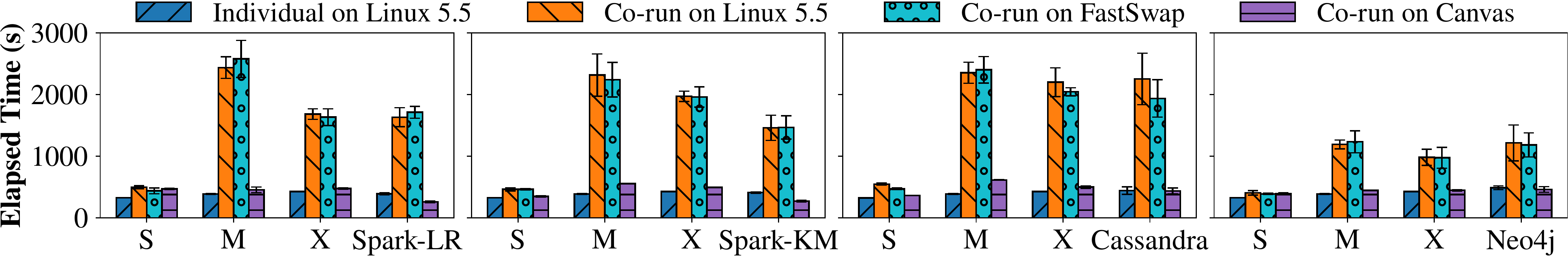} \\
          (a) 25\% local memory. \\
        \hspace{-0.5em} \includegraphics[scale=.4]{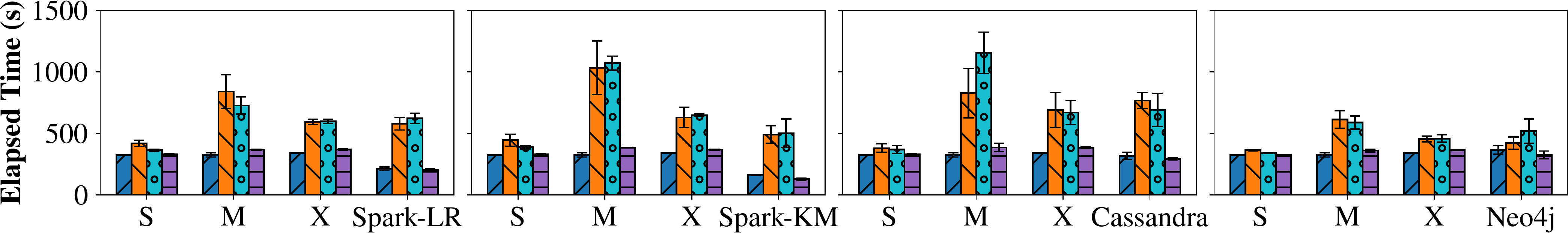} \\
          (b) 50\% local memory.
      \end{tabular}
    \caption{Performance of each program under 25\% and 50\% local memory when the three native programs, Snappy (S), Memcached (M), and XGBoost (X), co-run with a managed application. \tool ran with all optimizations enabled.} \label{fig:eval-overall}
    \vspace{-0.5em}
\end{figure*}

\mysubsection{Basic Swap Systems\label{sec:compare_baseline_swap}}

We used Fastswap~\cite{fastswap-eurosys20} as our underlying swap system, with a small amount of code changes to port Fastswap (originally built against Linux 4.11) to Linux 5.5. We first compared the performance of each individual application running on basic swap systems including Infiniswap~\cite{infiniswap-nsdi17}, Infiniswap with Leap~\cite{leap-atc20}, the original Fastswap~\cite{fastswap-eurosys20}, and \tool's ported Fastswap (without isolation and optimizations). We could not run LegoOS~\cite{legoos-osdi18} as it does not support network-related system calls, which are required for applications such as Spark. LegoOS implements swaps with RPCs as opposed to paging, but our idea (\ie, isolation and adaptive swapping) is applicable to this approach as well.

We ran Infiniswap and Leap on Linux 4.4, and Fastswap on Linux 4.11. The results are reported in Figure~\ref{fig:baseline-swap-systems}.
Infiniswap hung on XGBoost and Spark, and its corresponding bars are thus not reported in Figure~\ref{fig:baseline-swap-systems}.
Since \tool-swap was built off Fastswap, they have similar performance.

\mysubsection{Overall Performance \label{sec:eval_overall}}

Next, we demonstrate the overall performance when applications co-run together under \tool. Each experiment ran the same set of three native programs with one managed application: Spark-LR, Spark-KM, Cassandra, or Neo4j.
The results for the 25\% and 50\% local memory configurations are reported in Figure~\ref{fig:eval-overall}(a) and (b), respectively.

The four bars in each group represent an application's performance when running alone on Linux 5.5, co-running with other applications on Linux 5.5, co-running on the original Fastswap, and co-running on \tool (with all optimizations enabled). Across all experiments, \tool improves applications' co-run performance by up to \textbf{6.2$\times$} (average \textbf{3.5$\times$}) and up to \textbf{3.8$\times$} (average \textbf{1.9$\times$}) under the two memory configurations.
\tool enables Spark and Neo4j to even outperform their individual runs due to the optimizations that could also improve single-application performance.

\mysubsection{Isolation Reduces Degradation and Variation \label{sec:eval_isolation}}

\begin{figure*} [t]
    \centering
    \includegraphics[scale=.4]{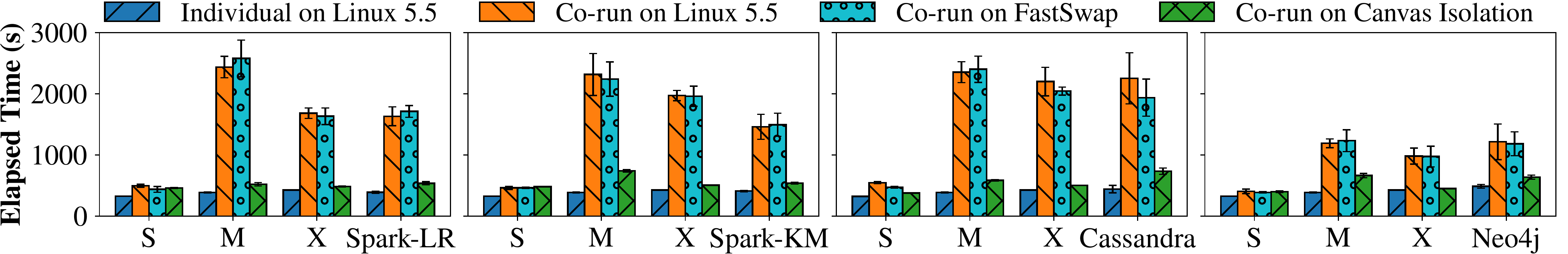}
    \begin{tabular}{cccc}
        (a) Co-run with Spark-LR. \label{fig:slr-isolation} &
        (b) Co-run with Spark-KM. \label{fig:skm-isolation} &
        (c) Co-run with Cassandra. \label{fig:cas-isolation} &
        (d) Co-run with Neo4j. \label{fig:n4j-isolation}
    \end{tabular}
    \vspace{-0.6em}
    \caption{Performance of native applications co-run with different managed applications under 25\% local memory; for \tool, only isolation was enabled (\ie, without adaptive optimizations).
    \label{fig:isolation-performance}}
    \vspace{-1.0em}
\end{figure*}

This experiment measures the effectiveness of isolation alone. We used a variant of \tool with the isolated swap system and RDMA bandwidth (\ie, vertical scheduling between applications) but without our swap-entry optimization, two-tier prefetcher, and horizontal RDMA scheduling.

\MyPara{Degradation Reduction.} We ran the same set of experiments under 25\% local memory. As shown in Figure~\ref{fig:isolation-performance},
isolation reduces the running time by up to 5.2$\times$, with an average of 2.5$\times$.
Isolation is particularly useful for applications that do not have many threads but need to frequently access remote memory, such as Memcached, which has 4 threads and
cannot compete for resources with managed applications such as Spark and Cassandra, which have more than 90 (application and runtime) threads.
As such, its performance is improved by \textbf{3.3$\times$} with dedicated swap resources. Isolation improves the average RDMA utilization by 2.8$\times$ from 692MB/s to 1908MB/s, making the peak bandwidth reach 4494MB/s.

\begin{table}[!htp]\centering
\caption{Performance variations of three native applications when co-running with each of the 11 managed applications under 25\% local memory (\tool{}~/ Linux 5.5~/ Fastswap).\label{tab:native-predictable}}
\scriptsize
\begin{adjustbox}{max width=\linewidth}
\begin{tabular}{l|rrr|rrr|rrr|rrr}\toprule
Program &\multicolumn{3}{c|}{Mean} &\multicolumn{3}{c|}{Min} &\multicolumn{3}{c|}{Max} &\multicolumn{3}{c}{$\sigma$} \\\midrule
Snappy &\textbf{1.07} &1.28 &1.23 &\textbf{1.03} &1.10 &1.08 &\textbf{1.23} &1.69 &1.46 &\textbf{0.07} &0.20 &0.14 \\
Memcached &\textbf{1.45} &3.24 &3.76 &\textbf{1.30} &1.48 &2.05 &\textbf{1.91} &6.05 &8.17 &\textbf{0.20} &1.82 &2.14 \\
XGBoost &\textbf{1.05} &3.17 &2.81 &\textbf{1.01} &1.38 &1.91 &\textbf{1.13} &6.13 &4.76 &\textbf{0.04} &1.59 &1.11 \\
Overall &\textbf{1.21} &2.56 &2.60 &\textbf{1.01} &1.10 &1.08 &\textbf{1.91} &6.13 &8.17 &\textbf{0.23} &1.64 &1.72 \\
\bottomrule
\end{tabular}
\end{adjustbox}
\end{table}

\MyPara{Variation Reduction. } One significant impact of interference is performance variation\textemdash the same application has drastically different performance when co-running with different applications (as shown in Figure~\ref{fig:multi-app-time}). To demonstrate our benefits, we co-ran the three native applications with each of the eleven managed applications listed in Table~\ref{tab:spark_programs}, which cover a wide spectrum of computation and memory access behaviors. Table~\ref{tab:native-predictable} reports various statistics of their performance including the mean, minimum, maximum, and standard deviation of their slowdowns (compared to their individual runs). Clearly, the performance of the three programs is much more stable (indicated by a small $\sigma$) under \tool than Linux\textemdash variations are reduced by 7$\times$ overall.

\mysubsection{Effectiveness of Adaptive Optimizations \label{sec:eval_optimization}}

\begin{figure} [t]
    \centering
    \includegraphics[scale=.39]{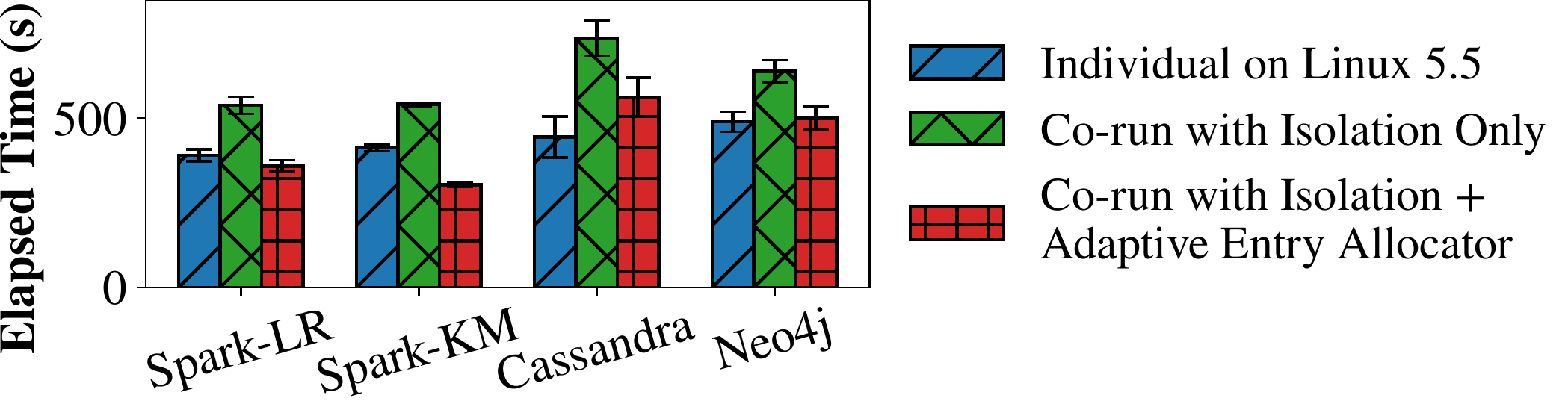}
    \caption{Benefit of adaptive swap entry allocation. Compared are the times of the application running individually on Linux 5.5, co-running on \tool with adaptive entry allocation disabled, and enabled.} \label{fig:adaptive-allocation}
    \vspace{-1.0em}
\end{figure}

This subsection evaluates the benefit of each swap optimization \emph{on top of the isolated swap system} by turning it on/off.

\mysubsubsection{Adaptive Swap Entry Allocator~\label{sec:swap_optimization}}
Isolation already reduces lock contention at swap entry allocation because each process has its own swap entry manager.
However, for multi-threaded applications such as Spark and Cassandra, their processing threads still have to go through the locking process. In this subsection, we focus on managed applications due to their extensive use of threads.
Figure~\ref{fig:adaptive-allocation} shows the performance of Spark LR, Spark KM, Cassandra, and Neo4j when they each co-run with the other three native programs.
On average, our adaptive allocation enables an \emph{additional} boost of 1.50$\times$ for Spark LR, 1.77$\times$ for Spark KM, 1.31$\times$ for Cassandra, and 1.28$\times$ for Neo4j.

\begin{table}[!htp]\centering
\vspace{-.3em}
\caption{Swap-out throughput w/ and w/o adaptive swap-entry allocation when native programs co-run with Spark.  \label{tab:benefits_of_adaptive_swap_entry_allocation}}
\begin{adjustbox}{max width=\linewidth}
\begin{tabular}{lrrrr}\toprule
Thruput (KPages/s) & Linux 5.5  & \tool w/o adap. alloc.  &  \tool w/  \\\midrule
Avg. Spark apps & 98 & 164 & \textbf{295} \\
Avg. all apps & 185 & 309 & \textbf{468} \\
\bottomrule
\end{tabular}
\end{adjustbox}
\end{table}

Table~\ref{tab:benefits_of_adaptive_swap_entry_allocation} reports the swap-out throughput when the native applications co-run with Spark. As shown, isolation improves the throughput by $1.67\times$ while adaptive allocation provides an additional boost of $1.51\times$. This benefit is obtained after applying all optimizations in Linux 5.5.

\begin{figure} [h!]
    \centering
    \begin{tabular}{ll}
     \includegraphics[scale=.36]{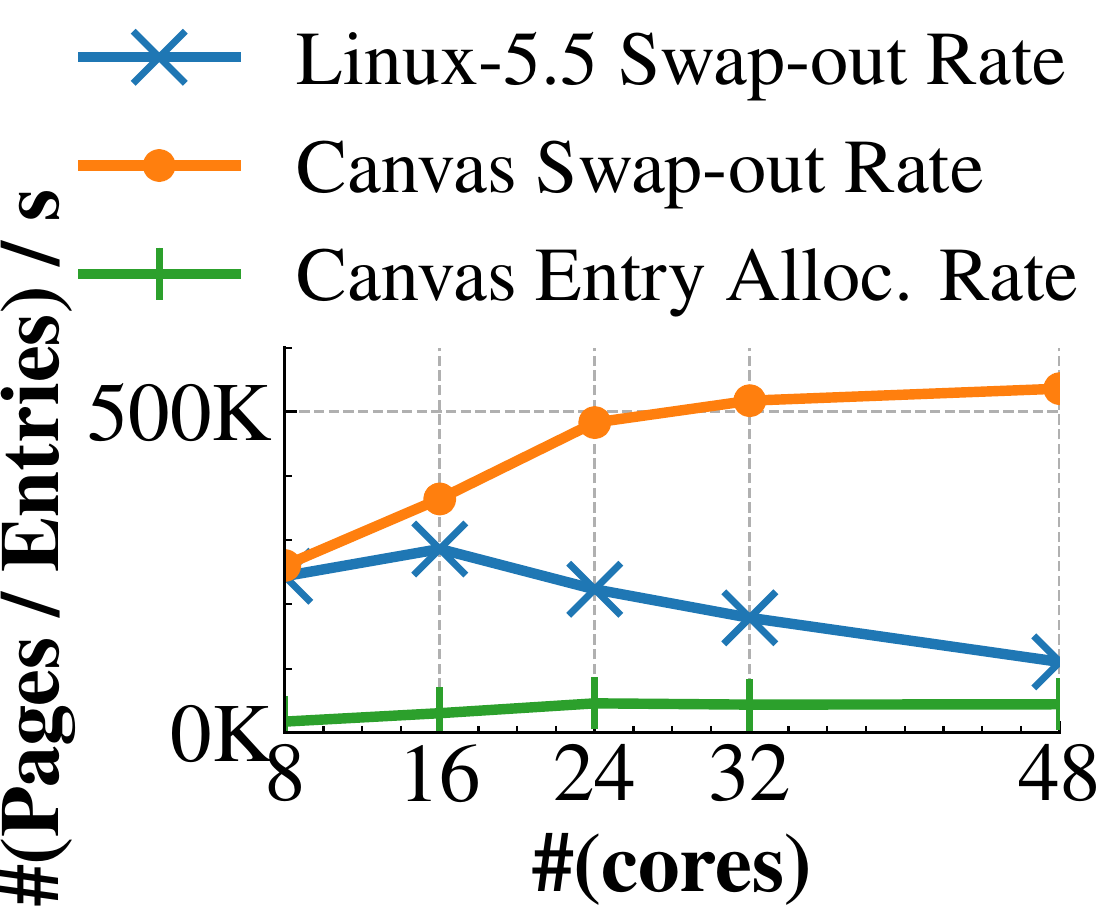} \label{fig:memcached-alloc-rate}
     &
     \hspace{-3em}
      \includegraphics[scale=.36]{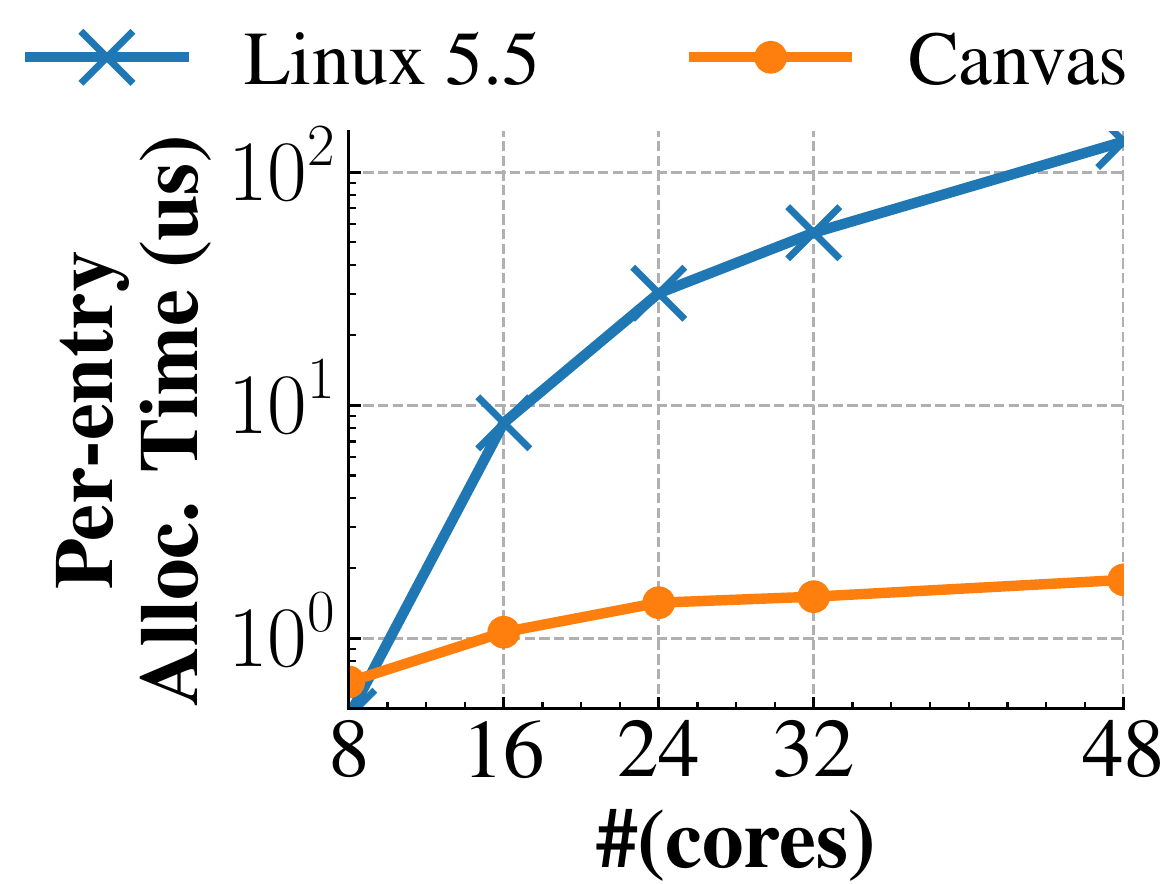} \label{fig:memcached-alloc-perf}\\
    (a) Swap-out and entry alloc rates. & (b) Per-entry alloc time.
    \end{tabular}
    \caption{Entry allocation comparison between the allocation algorithm in \tool and Linux 5.5 for Memcached under 25\% local memory. The Y-axis in (b) is log-scaled.\label{fig:alloc-memcached}}
    \vspace{-1.2em}
\end{figure}

\MyPara{Effectiveness of Entry Reservation.}
We compared our adaptive allocation algorithm with the original allocator in Linux 5.5 by running Memcached with varying (8 -- 48) cores under 25\% local memory.
As shown in Figure~\ref{fig:alloc-memcached}(a), for \tool, (1) the swap-out rate increases with the core number (showing good scalability) and (2) the swap entry allocation rate remains low. This is due to \tool's entry reservation algorithm that effectively reuses a significant number of swap entries for page swap-outs.
On the contrary, in Linux 5.5, the swap-out rate (which is the same as its entry allocation rate) decreases when more cores are used. This is because each entry allocation takes significantly longer, reducing the swap-out throughput. A comparison of per-entry allocation time can be seen in Figure~\ref{fig:alloc-memcached}(b).
We additionally compared the allocation algorithm between \tool, Linux 5.5, and Linux 5.14; these results are reported in Appendix \ref{sec:kernelpatches}.

\mysubsubsection{Prefetching Effectiveness}

Our baseline is the kernel's default prefetcher on the isolated swap system with adaptive swap allocator \emph{enabled}.
Since application-tier prefetching is designed primarily for high-level languages, here we focus on managed programs.

\MyPara{Time.} We compare the running time for three Spark applications LR, KM, TC, and Neo4j, between the kernel's prefetcher over \tool's isolated swap system and \tool's two-tier prefetcher, when each managed application co-runs with the three native applications under the 25\% local memory configuration. Application-tier prefetching brings \textbf{33\%}, \textbf{17\%}, \textbf{19\%}, and \textbf{8\%} additional performance benefits on top of the kernel prefetching with the isolated swap system. All the four managed applications benefit from the thread-level pattern analysis while the managed applications have seen 5-9\% contributions from using the reference-based pattern. The thread-level pattern analysis we added for native programs brings a 5\% and 11\% improvement for Memcached and XGBoost.%

We have also run Leap~\cite{leap-atc20}, a prefetcher that aggressively prefetches a number of contiguous pages if it cannot find any pattern. This approach may work for native applications because these applications access arrays; hence, the contiguous pages aggressively prefetched are likely to be useful for array accesses.  However, it works poorly for high-level language applications such as Spark and Neo4j, which use deep data structures and run graph-traversal GC tasks (which exhibit neither sequential nor strided patterns). Aggressively prefetching useless pages wastes the RDMA bandwidth and the swap cache.
Leap slows down our managed applications by 1.4$\times$, compared to the kernel's default prefetcher.

\begin{table}[!htp]\centering
\caption{Prefetching contribution and accuracy when different Spark and Neo4j co-run with native applications.}\label{tab:uffd-contribution}
\small
\begin{tabular}{lcccc}\toprule
Contribution &Spark-LR &Spark-KM &Spark-TC & Neo4j\\\midrule
Leap &23.4\% &25.8\% &42.2\% & \textbf{67.0\%} \\
Kernel &63.3\% &68.0\% & 65.9\% & 41.1\% \\
Canvas Two-tier &\textbf{79.2\%} &\textbf{79.3\%} & \textbf{75.3\%} & 45.0\% \\\midrule
Accuracy &Spark-LR &Spark-KM &Spark-TC & Neo4j\\\midrule
Leap &16.8\% &17.2\% &35.9\% & 6.1\%\\
Kernel &95.6\% &96.4\% & 93.9\% & 80.4\% \\
Canvas Two-tier &\textbf{94.3\%} &\textbf{94.8\%} &\textbf{94.9\%} &\textbf{87.1\%} \\
\bottomrule
\end{tabular}
\end{table}

\MyPara{Prefetching Contribution and Accuracy.} Table~\ref{tab:uffd-contribution} compares prefetching \emph{contribution} and \emph{accuracy} for the four managed applications when each of them co-runs with the same three native applications. Contribution is defined as a ratio between the number of page faults hitting on the swap cache and the total number of page faults (including both cache hits and demand swap-ins). Accuracy is defined as a ratio between the number of page faults hitting on the swap cache and the total number of prefetches. Clearly, contribution has a strong correlation with performance while accuracy measures the pattern recognition ability of a prefetcher. For example, for a conservative prefetcher that prefetches pages only if a pattern can be clearly identified, it can have a high accuracy (\ie, prefetched pages are all useful) but a low contribution (\ie, the number of prefetches is small).

Here we report prefetching contribution and accuracy for three prefetchers: Leap (on our isolated swap system), the kernel prefetcher (also on our isolated swap system),  and \tool's two-tier prefetcher. Among the three prefetchers,
for all but Neo4j, Leap has the lowest accuracy and contribution because it is an aggressive prefetcher. Leap keeps prefetching pages even when it cannot detect any patterns, which greatly reduces the prefetching accuracy. Second, due to the limited swap cache, the useless pages prefetched can cause previously prefetched pages to be released before they are accessed, hurting contribution. The kernel prefetcher and \tool have comparable accuracy because the kernel prefetcher is much more conservative than Leap. It stops prefetching when no clear pattern can be observed.  However, Linux has lower contribution than our two-tier prefetcher since \tool prefetches more useful pages using semantics.

\mysubsubsection{RDMA Scheduling}
We evaluate our two-dimensional RDMA scheduling. For the vertical dimension, we use the weighted min-max ratio (WMMR) $\frac{\min(x_i/w_i)}{\max(x_i/w_i)}$~\cite{shue-osdi12} as our bandwidth fairness metric (the closer to 1, the better),
where $x_i$ is the bandwidth consumption of the application $i$, and $w_i$ is its weight.
We set the weight proportionally to the average bandwidth of each application when running individually. Our vertical scheduling achieves an overall of \textbf{0.88} WMMR. %

\begin{figure}[h!]
   \begin{tabular}{cr}
\begin{minipage}[t]{.49\linewidth}
    \includegraphics[scale=0.37]{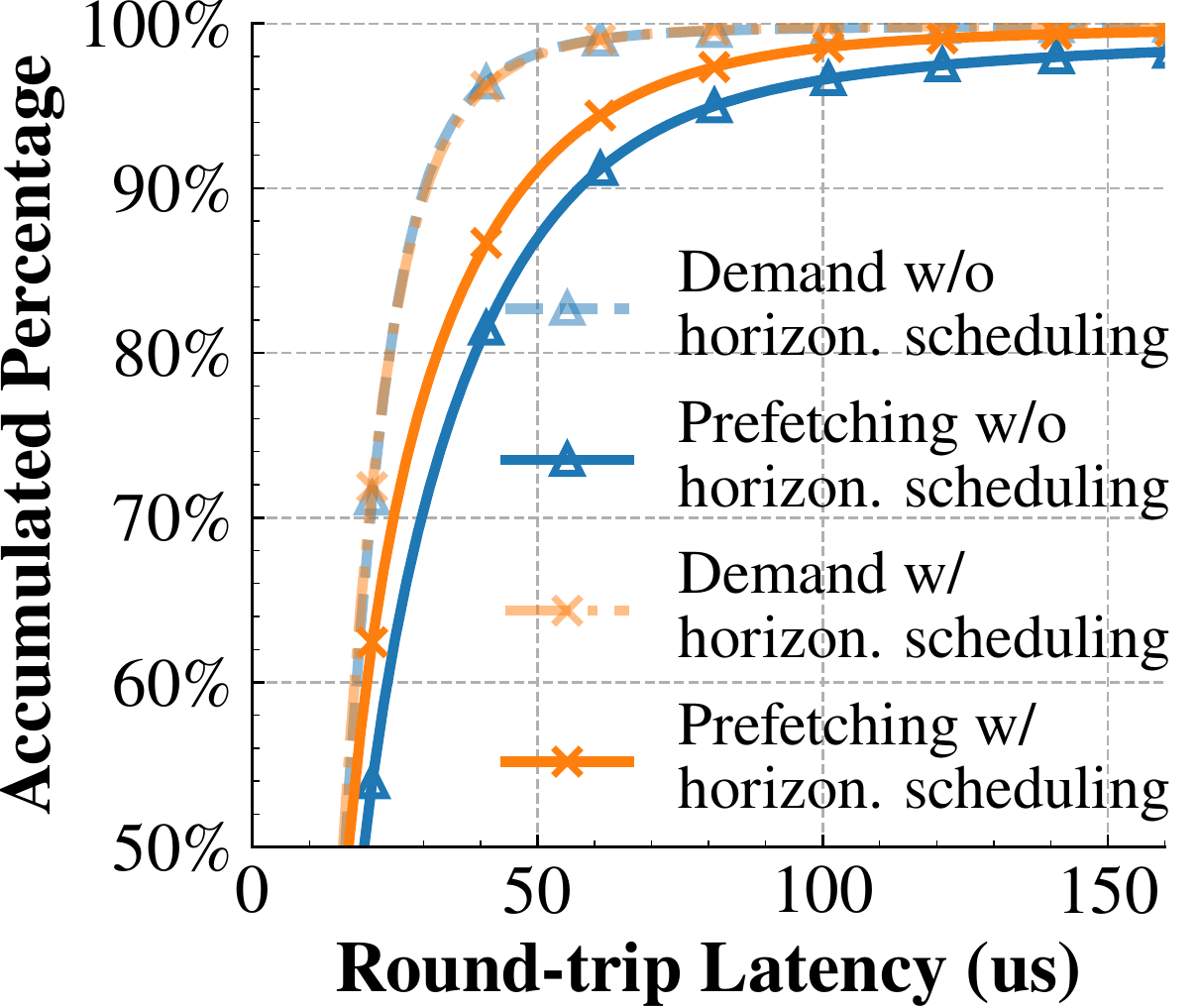}

\end{minipage}
&
\begin{minipage}[t]{.42\linewidth}
\vspace{-9.5em}
\small
\begin{tabular}{l}
\toprule
w/o horizontal scheduling  \\
Contribution: 52.1\%\\
Accuracy: 46.9\% \\\midrule
w/ horizontal scheduling   \\
Contribution: \textbf{62.8\%} \\
Accuracy: \textbf{52.4\%} \\
\bottomrule
\end{tabular}
\vspace{-1em}
\end{minipage}\\
(a) Latency CDF. & \hspace{-0.6em} (b) Prefetching effectiveness.\\
\end{tabular}
\caption{Horizontal scheduling effectiveness for GraphX-CC: (a) prefetching latency reduced, and (b) prefetching contribution and accuracy improved.  \label{fig:scheduler-latency}}
\vspace{-0.5em}
\end{figure}

The horizontal dimension (\ie, priority scheduling with timeliness) is our focus here because interference between prefetching and demand swapping is a unique challenge we overcome in this work. We ran GraphX Connected Components (GraphX-CC) with the three native applications. Figure~\ref{fig:scheduler-latency} compares the latency of sync vs.~async swap-in requests with and without the horizontal scheduling of RDMA.

As shown, our scheduler does \emph{not} incur overhead for the synchronous, demand requests but reduces the (90th percentile) latency of the asynchronous prefetching requests by $\thicksim$5\%. Note that these results were obtained with \tool's two-tier prefetcher enabled, which already generates precise prefetching requests. With the Leap prefetcher, the (90th percentile) latency reduction can be as high as \textbf{9$\times$}. To understand how the latency reduction improves prefetching effectiveness, we have also compared the prefetching contribution and accuracy with and without the horizontal scheduling, as shown in Figure~\ref{fig:scheduler-latency}(b). Due to the high timeliness requirement of prefetching requests, even 5\% latency reduction can lead to noticeable improvements in prefetching\textemdash \eg, the contribution/accuracy of GraphX-CC increases by \textbf{10.7\%} and \textbf{5.5\%} on top of the two-tier prefetcher\textemdash which ultimately translate to a \textbf{7-12\%} overall improvement.%

\mysection{Related Work~\label{sec:rw}}
\vspace{-0.1em}
\MyPara{Remote Memory.}
The past few years have seen a proliferation of remote-memory systems
that built on the kernel's swap mechanisms (including recent works such as LegoOS~\cite{legoos-osdi18}, Infiniswap~\cite{infiniswap-nsdi17}, Fastswap~\cite{fastswap-eurosys20}, and Semeru~\cite{semeru@osdi2020} as well as earlier attempts~\cite{remote_mem_as_swap@cse91, remote-regions-atc18, remote_mem@sosp95,network_ramdisk@cluster_computing99, remote_mem_as_swap@osdi94,remote_mem_as_swap@cmpcon93, remote_mem_as_swap@hpdc99, memliner-osdi22}). %
Remote memory is part of a general trend of resource disaggregation in datacenters~\cite{disaggregated_mem@hotnets13,carnonari-hotnets17, resource-disaggregation-osdi16, disaggregated_mem@isca11,memory_wall_par_cm,disaggregated_mem@hpca12, memory-disaggregation-isca09,the-machine-ross15, far-memory-hotos19, angel-hotcloud20, shenango-nsdi19, shoal-nsdi19}, which holds the promise of improving resource utilization and simplifying new hardware adoption. Under disaggregated memory, application data are stored on memory servers, making swap interference a more serious problem.

\MyPara{Resource Isolation.} Interference exists in a wide variety of settings~\cite{bolt-asplos17,heracles-tocs16,hadoop-scheduling} and resource isolation is crucial for delivering reliable performance for user workloads.
There is a large body of work on isolation of various kinds of resources including compute time~\cite{li-ppopp09, bartolini-taco14, cherkasova-sigmetrics07}, processor caches~\cite{ginseng-atc16, ubik-asplos14, rebudget-asplos16}, memory bandwidth~\cite{memory-partition-pact12, memory-partitioning-isca14,pard-asplos15,memory-isolation-sigmetric07,bubble-flux-isca13}, I/O bandwidth~\cite{mclock-osdi10, shue-osdi12, io-isolation-osdi14, stout-atc10, ioflow-sosp13, argon-fast07, split-level-io-sosp15}, network bandwidth~\cite{ballani-sigcomm11, secondnet-co-next10, ghodsi-nsdi11,seawall-hotcloud10,faircloud-sigcomm12,silverline-hotcloud11,eyeq-hotcloud12}, congestion control~\cite{vcc-sigcomm16,acdc-tcp-sigcomm16}, as well as CPU involved in network processing~\cite{iron-nsdi18}. Techniques such as IX~\cite{ix-osdi14} and MTCP~\cite{mtcp-nsdi14} isolate data-plane and application processing at the core granularity.

\MyPara{Prefetching.} Prefetching has been extensively studied, in the design of hardware cache ~\cite{prefetch-hardware1,prefetch-hardware2,prefetch-hardware3,prefetch-hardware4, prefetch-survey}, compilers~\cite{prefetching-hardware5, lattner-pldi05,rabbah-asplos04,peled-isca15,kolli-micro13,ferdman-micro11}, as well as operating systems~\cite{prefetching-os, leap-atc20}. Detecting spatial patterns~\cite{file-system-unix} is a common way to prefetch data. For example, various hardware techniques~\cite{sherwood-micro00, joseph-isca97, akanksha-micro13} have been developed to identify patterns (\ie, sequential or stride) in addresses accessed.
Leap~\cite{leap-atc20} is a kernel prefetcher designed specifically for applications using remote memory.
Swap interference can reduce the effectiveness of any existing prefetchers, let alone that none of them consider complex (semantic) patterns.
Early work such as \cite{informed-prefetching-sosp95,cao-tocs96} proposes application-level prefetching for efficient file operations on slow disks. Our prefetcher is, however, designed for a new setting where applications trigger page faults frequently and read pages from fast remote memory, with much tighter latency budgets.

\MyPara{RDMA Optimizations.} There is a body of recent work on RDMA scheduling~\cite{qiu-apnet18,shen-infocomm20} and scalability improvement~\cite{lite-rdma-sosp17, chen-eurosys19, fasst-osdi16, kalia-atc16, justitia-nsdi22}. These techniques focus more on scalability when RDMA NICs are shared among multiple clients.

\mysection{Conclusion}

We observed swap resources must be isolated when multiple applications use remote memory simultaneously. As such, \tool isolates swap cache, swap partition, and RDMA bandwidth to prevent applications from invading each other's resources. Now that resource accounting is done separately for applications, \tool offers three optimizations that adapt kernel operations such as swap-entry allocation, prefetching, and RDMA scheduling to each application's resource usage, providing additional performance boosts.

\section*{Acknowledgement}
We thank the anonymous reviewers for
their valuable and thorough comments. We are
grateful to our shepherd Danyang Zhuo for his feedback.
This work is supported by NSF grants CNS-1703598, CCF-1723773, CNS-1763172, CCF-1764077, CNS-1907352, CHS-1956322, CNS-2007737, CNS-2006437, CNS-2128653, CCF-2106404, CNS-2106838,  CNS-2147909,  CNS-2152313, CNS-2151630, and CNS-2140552, CNS-2153449, ONR grant N00014-18-1-2037, a Sloan Research Fellowship, and research grants from Cisco, Intel CAPA, and Samsung.
\bibliographystyle{abbrv}
\bibliography{references}

\begin{thebibliography}{100}

\bibitem{compress-enwik9-ds}
Large {Text} {Compression} {Benchmark}.

\bibitem{nvmeof}
{NVMe} over fabrics.
\newblock
  \url{http://community.mellanox.com/s/article/what-is-nvme-over-fabrics-x}.

\bibitem{kdddata}
Libsvm data: Classification.
\newblock \url{https://www.csie.ntu.edu.tw/~cjlin/libsvmtools/datasets}, 2012.

\bibitem{memcached}
Memcached - a distributed memory object caching system.
\newblock http://memcached.org, 2020.

\bibitem{wikipedia-ds}
Konect networks data.
\newblock \url{http://konect.cc/networks/}, 2021.

\bibitem{remote-regions-atc18}
M.~K. Aguilera, N.~Amit, I.~Calciu, X.~Deguillard, J.~Gandhi, S.~Novakovic,
  A.~Ramanathan, P.~Subrahmanyam, L.~Suresh, K.~Tati, R.~Venkatasubramanian,
  and M.~Wei.
\newblock Remote regions: A simple abstraction for remote memory.
\newblock In {\em USENIX ATC}, pages 775--787, 2018.

\bibitem{far-memory-hotos19}
M.~K. Aguilera, K.~Keeton, S.~Novakovic, and S.~Singhal.
\newblock Designing far memory data structures: Think outside the box.
\newblock In {\em HotOS}, pages 120--126, 2019.

\bibitem{fastswap-eurosys20}
E.~Amaro, C.~Branner-Augmon, Z.~Luo, A.~Ousterhout, M.~K. Aguilera, A.~Panda,
  S.~Ratnasamy, and S.~Shenker.
\newblock Can far memory improve job throughput?
\newblock In {\em EuroSys}, 2020.

\bibitem{angel-hotcloud20}
S.~Angel, M.~Nanavati, and S.~Sen.
\newblock Disaggregation and the application.
\newblock In {\em HotCloud}, 2020.

\bibitem{cassandra}
Apache.
\newblock Apache cassandra.
\newblock \url{https://cassandra.apache.org}, 2021.

\bibitem{memory_wall_par_cm}
K.~Asanović, R.~Bodik, B.~C. Catanzaro, J.~J. Gebis, P.~Husbands, K.~Keutzer,
  D.~A. Patterson, W.~L. Plishker, J.~Shalf, S.~W. Williams, and K.~A. Yelick.
\newblock The landscape of parallel computing research: A view from berkeley.
\newblock Technical Report UCB/EECS-2006-183, EECS Department, University of
  California, Berkeley, Dec 2006.

\bibitem{higgs-ds}
P.~Baldi, P.~Sadowski, and D.~Whiteson.
\newblock Searching for exotic particles in high-energy physics with deep
  learning.
\newblock {\em Nature communications}, 5(1):1--9, 2014.

\bibitem{ballani-sigcomm11}
H.~Ballani, P.~Costa, T.~Karagiannis, and A.~Rowstron.
\newblock Towards predictable datacenter networks.
\newblock In {\em SIGCOMM}, pages 242--253, 2011.

\bibitem{disaggregated_mem@isca11}
L.~A. Barroso.
\newblock Warehouse-scale computing: Entering the teenage decade.
\newblock In {\em ISCA}, 2011.

\bibitem{datacenter-google}
L.~A. Barroso, U.~Hölzle, and P.~Ranganathan.
\newblock {\em The Datacenter as a Computer: Designing Warehouse-Scale
  Machines, Third Edition}.
\newblock Synthesis Lectures on Computer Architecture, 2018.

\bibitem{bartolini-taco14}
D.~B. Bartolini, F.~Sironi, D.~Sciuto, and M.~D. Santambrogio.
\newblock Automated fine-grained cpu provisioning for virtual machines.
\newblock {\em ACM Trans. Archit. Code Optim.}, 11(3), July 2014.

\bibitem{ix-osdi14}
A.~Belay, G.~Prekas, A.~Klimovic, S.~Grossman, C.~Kozyrakis, and E.~Bugnion.
\newblock {IX}: A protected dataplane operating system for high throughput and
  low latency.
\newblock In {\em OSDI}, pages 49--65, 2014.

\bibitem{max-min-fairness}
D.~Bertsekas and R.~Gallager.
\newblock {\em Data Networks (2nd Ed.)}.
\newblock Prentice-Hall, Inc., USA, 1992.

\bibitem{disaggregated-runtime-asplos21}
I.~Calciu, M.~T. Imran, I.~Puddu, S.~Kashyap, H.~A. Maruf, O.~Mutlu, and
  A.~Kolli.
\newblock Rethinking software runtimes for disaggregated memory.
\newblock In {\em ASPLOS}, pages 79--92, 2021.

\bibitem{cao-tocs96}
P.~Cao, E.~W. Felten, A.~R. Karlin, and K.~Li.
\newblock Implementation and performance of integrated application-controlled
  file caching, prefetching, and disk scheduling.
\newblock {\em ACM Trans. Comput. Syst.}, 14(4):311--343, Nov. 1996.

\bibitem{carnonari-hotnets17}
A.~Carbonari and I.~Beschasnikh.
\newblock Tolerating faults in disaggregated datacenters.
\newblock In {\em HotNets-XVI}, pages 164--170, 2017.

\bibitem{chen-kdd16}
T.~Chen and C.~Guestrin.
\newblock {XGBoost}: A scalable tree boosting system.
\newblock In {\em KDD}, pages 785--794, 2016.

\bibitem{xgboost}
T.~Chen and C.~Guestrin.
\newblock extreme gradient boosting for applied machine learning.
\newblock https://xgboost.readthedocs.io/en/latest/, 2021.

\bibitem{chen-eurosys19}
Y.~Chen, Y.~Lu, and J.~Shu.
\newblock Scalable {RDMA RPC} on reliable connection with efficient resource
  sharing.
\newblock In {\em EuroSys}, 2019.

\bibitem{cherkasova-sigmetrics07}
L.~Cherkasova, D.~Gupta, and A.~Vahdat.
\newblock Comparison of the three cpu schedulers in xen.
\newblock {\em SIGMETRICS Perform. Eval. Rev.}, 35(2):42--51, 2007.

\bibitem{cooper-ycsb-ds}
B.~F. Cooper, A.~Silberstein, E.~Tam, R.~Ramakrishnan, and R.~Sears.
\newblock Benchmarking cloud serving systems with ycsb.
\newblock In {\em Proceedings of the 1st ACM Symposium on Cloud Computing},
  SoCC '10, page 143–154, New York, NY, USA, 2010. Association for Computing
  Machinery.

\bibitem{vcc-sigcomm16}
B.~Cronkite-Ratcliff, A.~Bergman, S.~Vargaftik, M.~Ravi, N.~McKeown,
  I.~Abraham, and I.~Keslassy.
\newblock Virtualized congestion control.
\newblock In {\em SIGCOMM}, pages 230--243, 2016.

\bibitem{remote_mem_as_swap@osdi94}
M.~D. Dahlin, R.~Y. Wang, T.~E. Anderson, and D.~A. Patterson.
\newblock Cooperative caching: Using remote client memory to improve file
  system performance.
\newblock In {\em OSDI}, 1994.

\bibitem{bolt-asplos17}
C.~Delimitrou and C.~Kozyrakis.
\newblock Bolt: I know what you did last summer... in the cloud.
\newblock In {\em ASPLOS}, pages 599--613, 2017.

\bibitem{wfq-demers-sigcomm89}
A.~Demers, S.~Keshav, and S.~Shenker.
\newblock Analysis and simulation of a fair queueing algorithm.
\newblock {\em SIGCOMM Comput. Commun. Rev.}, 19(4):1–12, Aug. 1989.

\bibitem{remote_mem@sosp95}
M.~J. Feeley, W.~E. Morgan, E.~P. Pighin, A.~R. Karlin, H.~M. Levy, and C.~A.
  Thekkath.
\newblock Implementing global memory management in a workstation cluster.
\newblock In {\em SOSP}, pages 201--212, 1995.

\bibitem{remote_mem_as_swap@cse91}
E.~Felten and J.~Zahorjan.
\newblock Issues in the implementation of a remote memory paging system.
\newblock In {\em University of Washington CSE TR CSE TR}, 1991.

\bibitem{ferdman-micro11}
M.~Ferdman, C.~Kaynak, and B.~Falsafi.
\newblock Proactive instruction fetch.
\newblock In {\em MICRO}, pages 152--162, 2011.

\bibitem{network_ramdisk@cluster_computing99}
M.~D. Flouris and E.~P. Markatos.
\newblock The network ramdisk: Using remote memory on heterogeneous nows.
\newblock {\em Cluster Computing}, 2(4), Dec 1999.

\bibitem{ginseng-atc16}
L.~Funaro, O.~A. Ben-Yehuda, and A.~Schuster.
\newblock Ginseng: Market-driven {LLC} allocation.
\newblock In {\em USENIX ATC}, pages 295--308, 2016.

\bibitem{resource-disaggregation-osdi16}
P.~X. Gao, A.~Narayan, S.~Karandikar, J.~Carreira, S.~Han, R.~Agarwal,
  S.~Ratnasamy, and S.~Shenker.
\newblock Network requirements for resource disaggregation.
\newblock In {\em OSDI}, pages 249--264, 2016.

\bibitem{ghodsi-nsdi11}
A.~Ghodsi, M.~Zaharia, B.~Hindman, A.~Konwinski, S.~Shenker, and I.~Stoica.
\newblock Dominant resource fairness: Fair allocation of multiple resource
  types.
\newblock In {\em NSDI}, pages 323--336, 2011.

\bibitem{snappy}
Google.
\newblock Google's fast compressor/decompressor.
\newblock https://github.com/google/snappy, 2020.

\bibitem{infiniswap-nsdi17}
J.~Gu, Y.~Lee, Y.~Zhang, M.~Chowdhury, and K.~G. Shin.
\newblock Efficient memory disaggregation with infiniswap.
\newblock In {\em NSDI}, pages 649--667, 2017.

\bibitem{mclock-osdi10}
A.~Gulati, A.~Merchant, and P.~J. Varman.
\newblock {MClock}: Handling throughput variability for hypervisor {IO}
  scheduling.
\newblock In {\em OSDI}, pages 437--450, 2010.

\bibitem{secondnet-co-next10}
C.~Guo, G.~Lu, H.~J. Wang, S.~Yang, C.~Kong, P.~Sun, W.~Wu, and Y.~Zhang.
\newblock {SecondNet}: A data center network virtualization architecture with
  bandwidth guarantees.
\newblock In {\em Co-NEXT}, 2010.

\bibitem{prefetch-hardware2}
Y.~Guo.
\newblock {\em Compiler-Assisted Hardware-Based Data Prefetching for next
  Generation Processors}.
\newblock PhD thesis, 2007.

\bibitem{disaggregated_mem@hotnets13}
S.~Han, N.~Egi, A.~Panda, S.~Ratnasamy, G.~Shi, and S.~Shenker.
\newblock Network support for resource disaggregation in next-generation
  datacenters.
\newblock In {\em HotNets}, pages 10:1--10:7, 2013.

\bibitem{acdc-tcp-sigcomm16}
K.~He, E.~Rozner, K.~Agarwal, Y.~J. Gu, W.~Felter, J.~Carter, and A.~Akella.
\newblock {AC/DC TCP}: Virtual congestion control enforcement for datacenter
  networks.
\newblock In {\em SIGCOMM}, pages 244--257, 2016.

\bibitem{remote_mem_as_swap@cmpcon93}
L.~{Iftode}, K.~{Li}, and K.~{Petersen}.
\newblock Memory servers for multicomputers.
\newblock In {\em Digest of Papers. Compcon Spring}, pages 538--547, Feb 1993.

\bibitem{patch2}
Intel.
\newblock Batch allocation for swap entries.
\newblock \url{https://github.com/torvalds/linux/commit/ed43af10975eef7e},
  2020.

\bibitem{patch3}
Intel.
\newblock Memcontrol: Charge swap-in pages to \texttt{cgroup}.
\newblock \url{https://github.com/torvalds/linux/commit/4c6355b25e8bb83c},
  2020.

\bibitem{patch1}
Intel.
\newblock Per-core cluster allocation.
\newblock \url{https://github.com/torvalds/linux/commit/490705888107c3ed},
  2020.

\bibitem{perfiso-atc18}
C.~Iorgulescu, R.~Azimi, Y.~Kwon, S.~Elnikety, M.~Syamala, V.~Narasayya,
  H.~Herodotou, P.~Tomita, A.~Chen, J.~Zhang, and J.~Wang.
\newblock {PerfIso}: Performance isolation for commercial {Latency-Sensitive}
  services.
\newblock In {\em USENIX ATC}, pages 519--532, 2018.

\bibitem{memory-isolation-sigmetric07}
R.~Iyer, L.~Zhao, F.~Guo, R.~Illikkal, S.~Makineni, D.~Newell, Y.~Solihin,
  L.~Hsu, and S.~Reinhardt.
\newblock {QoS} policies and architecture for cache/memory in {CMP} platforms.
\newblock In {\em SIGMETRICS}, pages 25--36, 2007.

\bibitem{akanksha-micro13}
A.~Jain and C.~Lin.
\newblock Linearizing irregular memory accesses for improved correlated
  prefetching.
\newblock In {\em MICRO}, pages 247--259, 2013.

\bibitem{mtcp-nsdi14}
E.~Y. Jeong, S.~Woo, M.~Jamshed, H.~Jeong, S.~Ihm, D.~Han, and K.~Park.
\newblock {MTCP}: A highly scalable user-level {TCP} stack for multicore
  systems.
\newblock In {\em NSDI}, pages 489--502, 2014.

\bibitem{eyeq-hotcloud12}
V.~Jeyakumar, M.~Alizadeh, D.~Mazi{\`e}res, B.~Prabhakar, and C.~Kim.
\newblock {EyeQ}: Practical network performance isolation for the multi-tenant
  cloud.
\newblock In {\em HotCloud}, 2012.

\bibitem{joseph-isca97}
D.~Joseph and D.~Grunwald.
\newblock Prefetching using markov predictors.
\newblock In {\em ISCA}, pages 252--263, 1997.

\bibitem{kalia-atc16}
A.~Kalia, M.~Kaminsky, and D.~G. Andersen.
\newblock Design guidelines for high performance {RDMA} systems.
\newblock In {\em USENIX ATC}, pages 437--450, 2016.

\bibitem{fasst-osdi16}
A.~Kalia, M.~Kaminsky, and D.~G. Andersen.
\newblock {FaSST}: Fast, scalable and simple distributed transactions with
  two-sided ({RDMA}) datagram {RPCs}.
\newblock In {\em OSDI}, pages 185--201, 2016.

\bibitem{ubik-asplos14}
H.~Kasture and D.~Sanchez.
\newblock {Ubik}: Efficient cache sharing with strict qos for latency-critical
  workloads.
\newblock In {\em ASPLOS}, pages 729--742, 2014.

\bibitem{the-machine-ross15}
K.~Keeton.
\newblock {The Machine}: An architecture for memory-centric computing.
\newblock In {\em ROSS}, 2015.

\bibitem{iron-nsdi18}
J.~Khalid, E.~Rozner, W.~Felter, C.~Xu, K.~Rajamani, A.~Ferreira, and
  A.~Akella.
\newblock Iron: Isolating network-based cpu in container environments.
\newblock In {\em NSDI}, pages 313--328, 2018.

\bibitem{kolli-micro13}
A.~{Kolli}, A.~{Saidi}, and T.~F. {Wenisch}.
\newblock {RDIP}: Return-address-stack directed instruction prefetching.
\newblock In {\em MICRO}, pages 260--271, 2013.

\bibitem{remote_mem_as_swap@hpdc99}
S.~{Koussih}, A.~{Acharya}, and S.~{Setia}.
\newblock Dodo: a user-level system for exploiting idle memory in workstation
  clusters.
\newblock In {\em HPDC}, pages 301--308, Aug 1999.

\bibitem{lagar-cavilla-asplos19}
A.~Lagar-Cavilla, J.~Ahn, S.~Souhlal, N.~Agarwal, R.~Burny, S.~Butt, J.~Chang,
  A.~Chaugule, N.~Deng, J.~Shahid, G.~Thelen, K.~A. Yurtsever, Y.~Zhao, and
  P.~Ranganathan.
\newblock Software-defined far memory in warehouse-scale computers.
\newblock In {\em ASPLOS}, pages 317--330, 2019.

\bibitem{lattner-pldi05}
C.~Lattner and V.~Adve.
\newblock Automatic pool allocation: improving performance by controlling data
  structure layout in the heap.
\newblock In {\em PLDI}, pages 129--142, 2005.

\bibitem{li-ppopp09}
T.~Li, D.~Baumberger, and S.~Hahn.
\newblock Efficient and scalable multiprocessor fair scheduling using
  distributed weighted round-robin.
\newblock In {\em PPoPP}, pages 65--74, 2009.

\bibitem{memory-disaggregation-isca09}
K.~Lim, J.~Chang, T.~Mudge, P.~Ranganathan, S.~K. Reinhardt, and T.~F. Wenisch.
\newblock Disaggregated memory for expansion and sharing in blade servers.
\newblock In {\em ISCA}, pages 267--278, 2009.

\bibitem{disaggregated_mem@hpca12}
K.~{Lim}, Y.~{Turner}, J.~R. {Santos}, A.~{AuYoung}, J.~{Chang},
  P.~{Ranganathan}, and T.~F. {Wenisch}.
\newblock System-level implications of disaggregated memory.
\newblock In {\em HPCA}, pages 1--12, 2012.

\bibitem{memory-partition-pact12}
L.~Liu, Z.~Cui, M.~Xing, Y.~Bao, M.~Chen, and C.~Wu.
\newblock A software memory partition approach for eliminating bank-level
  interference in multicore systems.
\newblock In {\em PACT}, pages 367--376, 2012.

\bibitem{memory-partitioning-isca14}
L.~Liu, Y.~Li, Z.~Cui, Y.~Bao, M.~Chen, and C.~Wu.
\newblock Going vertical in memory management: Handling multiplicity by
  multi-policy.
\newblock In {\em ISCA}, pages 169--180, 2014.

\bibitem{heracles-tocs16}
D.~Lo, L.~Cheng, R.~Govindaraju, P.~Ranganathan, and C.~Kozyrakis.
\newblock Improving resource efficiency at scale with heracles.
\newblock {\em ACM Trans. Comput. Syst.}, 34(2), 2016.

\bibitem{io-isolation-osdi14}
L.~Lu, Y.~Zhang, T.~Do, S.~Al-Kiswany, A.~C. Arpaci-Dusseau, and R.~H.
  Arpaci-Dusseau.
\newblock Physical disentanglement in a container-based file system.
\newblock In {\em OSDI}, pages 81--96, 2014.

\bibitem{pard-asplos15}
J.~Ma, X.~Sui, N.~Sun, Y.~Li, Z.~Yu, B.~Huang, T.~Xu, Z.~Yao, Y.~Chen, H.~Wang,
  L.~Zhang, and Y.~Bao.
\newblock Supporting differentiated services in computers via programmable
  architecture for resourcing-on-demand ({PARD}).
\newblock In {\em ASPLOS}, pages 131--143, 2015.

\bibitem{taurus-asplos16}
M.~Maas, K.~Asanovi\'{c}, T.~Harris, and J.~Kubiatowicz.
\newblock Taurus: A holistic language runtime system for coordinating
  distributed managed-language applications.
\newblock In {\em ASPLOS}, pages 457--471, 2016.

\bibitem{leap-atc20}
H.~A. Maruf and M.~Chowdhury.
\newblock Effectively prefetching remote memory with {L}eap.
\newblock In {\em USENIX ATC}, pages 843--857, 2020.

\bibitem{stout-atc10}
J.~C. McCullough, J.~Dunagan, A.~Wolman, and A.~C. Snoeren.
\newblock Stout: An adaptive interface to scalable cloud storage.
\newblock In {\em USENIX ATC}, 2010.

\bibitem{file-system-unix}
M.~K. McKusick, W.~N. Joy, S.~J. Leffler, and R.~S. Fabry.
\newblock A fast file system for {UNIX}.
\newblock {\em ACM Trans. Comput. Syst.}, 2(3):181--197, 1984.

\bibitem{prefetch-survey}
S.~Mittal.
\newblock A survey of recent prefetching techniques for processor caches.
\newblock {\em ACM Comput. Surv.}, 49(2), 2016.

\bibitem{silverline-hotcloud11}
Y.~Mundada, A.~Ramachandran, and N.~Feamster.
\newblock Silverline: Data and network isolation for cloud services.
\newblock In {\em HotCloud}, 2011.

\bibitem{gerenuk-sosp19}
C.~Navasca, C.~Cai, K.~Nguyen, B.~Demsky, S.~Lu, M.~Kim, and G.~H. Xu.
\newblock Gerenuk: Thin computation over big native data using speculative
  program transformation.
\newblock In {\em SOSP}, pages 538--553, 2019.

\bibitem{neo4j}
Neo4j.
\newblock Neo4j graph data platform.
\newblock \url{https://neo4j.com}, 2021.

\bibitem{skyway-asplos18}
K.~Nguyen, L.~Fang, C.~Navasca, G.~Xu, B.~Demsky, and S.~Lu.
\newblock Skyway: Connecting managed heaps in distributed big data systems.
\newblock In {\em ASPLOS}, pages 56--69, 2018.

\bibitem{yak-osdi16}
K.~Nguyen, L.~Fang, G.~Xu, B.~Demsky, S.~Lu, S.~Alamian, and O.~Mutlu.
\newblock Yak: A high-performance big-data-friendly garbage collector.
\newblock In {\em OSDI}, pages 349--365, 2016.

\bibitem{facade-asplos15}
K.~Nguyen, K.~Wang, Y.~Bu, L.~Fang, J.~Hu, and G.~Xu.
\newblock Facade: A compiler and runtime for (almost) object-bounded big data
  applications.
\newblock In {\em ASPLOS}, pages 675--690, 2015.

\bibitem{shenango-nsdi19}
A.~Ousterhout, J.~Fried, J.~Behrens, A.~Belay, and H.~Balakrishnan.
\newblock Shenango: Achieving high {CPU} efficiency for latency-sensitive
  datacenter workloads.
\newblock In {\em NSDI}, pages 361--378, 2019.

\bibitem{pgps-parekh-1993}
A.~Parekh and R.~Gallager.
\newblock A generalized processor sharing approach to flow control in
  integrated services networks: the single-node case.
\newblock {\em IEEE/ACM Transactions on Networking}, 1(3):344--357, 1993.

\bibitem{informed-prefetching-sosp95}
R.~H. Patterson, G.~A. Gibson, E.~Ginting, D.~Stodolsky, and J.~Zelenka.
\newblock Informed prefetching and caching.
\newblock In {\em SOSP}, pages 79--95, 1995.

\bibitem{peled-isca15}
L.~Peled, S.~Mannor, U.~Weiser, and Y.~Etsion.
\newblock Semantic locality and context-based prefetching using reinforcement
  learning.
\newblock In {\em ISCA}, pages 285--297, 2015.

\bibitem{faircloud-sigcomm12}
L.~Popa, A.~Krishnamurthy, S.~Ratnasamy, and I.~Stoica.
\newblock Faircloud: Sharing the network in cloud computing.
\newblock In {\em SIGCOMM}, pages 187--198, 2012.

\bibitem{qiu-apnet18}
H.~Qiu, X.~Wang, T.~Jin, Z.~Qian, B.~Ye, B.~Tang, W.~Li, and S.~Lu.
\newblock Toward effective and fair {RDMA} resource sharing.
\newblock In {\em APNet}, pages 8--14, 2018.

\bibitem{rabbah-asplos04}
R.~M. Rabbah, H.~Sandanagobalane, M.~Ekpanyapong, and W.-F. Wong.
\newblock Compiler orchestrated prefetching via speculation and predication.
\newblock In {\em ASPLOS}, pages 189--198, 2004.

\bibitem{aifm@osdi2020}
Z.~Ruan, M.~Schwarzkopf, M.~K. Aguilera, and A.~Belay.
\newblock {AIFM}: High-performance, application-integrated far memory.
\newblock In {\em OSDI}, pages 315--332, 2020.

\bibitem{legoos-osdi18}
Y.~Shan, Y.~Huang, Y.~Chen, and Y.~Zhang.
\newblock {LegoOS}: A disseminated, distributed {OS} for hardware resource
  disaggregation.
\newblock In {\em OSDI}, pages 69--87, 2018.

\bibitem{shen-infocomm20}
D.~Shen, J.~Luo, F.~Dong, X.~Guo, K.~Wang, and J.~C.~S. Lui.
\newblock Distributed and optimal rdma resource scheduling in shared data
  center networks.
\newblock In {\em INFOCOM}, pages 606--615, 2020.

\bibitem{sherwood-micro00}
T.~Sherwood, S.~Sair, and B.~Calder.
\newblock Predictor-directed stream buffers.
\newblock In {\em MICRO}, pages 42--53, 2000.

\bibitem{seawall-hotcloud10}
A.~Shieh, S.~Kandula, A.~Greenberg, and C.~Kim.
\newblock Seawall: Performance isolation for cloud datacenter networks.
\newblock In {\em HotCloud}, 2010.

\bibitem{shoal-nsdi19}
V.~Shrivastav, A.~Valadarsky, H.~Ballani, P.~Costa, K.~S. Lee, H.~Wang,
  R.~Agarwal, and H.~Weatherspoon.
\newblock Shoal: A network architecture for disaggregated racks.
\newblock In {\em NSDI}, pages 255--270, 2019.

\bibitem{shue-osdi12}
D.~Shue, M.~J. Freedman, and A.~Shaikh.
\newblock Performance isolation and fairness for multi-tenant cloud storage.
\newblock In {\em OSDI}, pages 349--362, 2012.

\bibitem{ioflow-sosp13}
E.~Thereska, H.~Ballani, G.~O'Shea, T.~Karagiannis, A.~Rowstron, T.~Talpey,
  R.~Black, and T.~Zhu.
\newblock {IOFlow}: A software-defined storage architecture.
\newblock In {\em SOSP}, pages 182--196, 2013.

\bibitem{prefetching-hardware5}
{Tien-Fu Chen} and {Jean-Loup Baer}.
\newblock Effective hardware-based data prefetching for high-performance
  processors.
\newblock {\em IEEE Transactions on Computers}, 44(5):609--623, 1995.

\bibitem{lite-rdma-sosp17}
S.-Y. Tsai and Y.~Zhang.
\newblock {LITE} kernel {RDMA} support for datacenter applications.
\newblock In {\em SOSP}, pages 306--324, 2017.

\bibitem{prefetch-hardware4}
S.~P. {Vander Wiel} and D.~J. {Lilja}.
\newblock When caches aren't enough: data prefetching techniques.
\newblock {\em Computer}, 30(7):23--30, 1997.

\bibitem{prefetch-hardware1}
S.~P. {Vander Wiel} and D.~J. {Lilja}.
\newblock A compiler-assisted data prefetch controller.
\newblock In {\em Proceedings 1999 IEEE International Conference on Computer
  Design: VLSI in Computers and Processors}, pages 372--377, 1999.

\bibitem{prefetching-os}
G.~M. Voelker, E.~J. Anderson, T.~Kimbrel, M.~J. Feeley, J.~S. Chase, A.~R.
  Karlin, and H.~M. Levy.
\newblock Implementing cooperative prefetching and caching in a
  globally-managed memory system.
\newblock In {\em SIGMETRICS}, pages 33--43, 1998.

\bibitem{argon-fast07}
M.~Wachs, M.~Abd-El-Malek, E.~Thereska, and G.~R. Ganger.
\newblock Argon: Performance insulation for shared storage servers.
\newblock In {\em FAST}, 2007.

\bibitem{semeru@osdi2020}
C.~Wang, H.~Ma, S.~Liu, Y.~Li, Z.~Ruan, K.~Nguyen, M.~D. Bond, R.~Netravali,
  M.~Kim, and G.~H. Xu.
\newblock Semeru: A memory-disaggregated managed runtime.
\newblock In {\em 14th {USENIX} Symposium on Operating Systems Design and
  Implementation ({OSDI} 20)}, pages 261--280. {USENIX} Association, Nov. 2020.

\bibitem{memliner-osdi22}
C.~Wang, H.~Ma, S.~Liu, Y.~Qiao, J.~Eyolfson, C.~Navasca, S.~Lu, and G.~H. Xu.
\newblock {MemLiner}: Lining up tracing and application for a
  {Far-Memory-Friendly} runtime.
\newblock In {\em OSDI}, pages 35--53, 2022.

\bibitem{rebudget-asplos16}
X.~Wang and J.~F. Mart\'{\i}nez.
\newblock {ReBudget}: Trading off efficiency vs. fairness in market-based
  multicore resource allocation via runtime budget reassignment.
\newblock In {\em ASPLOS}, pages 19--32, 2016.

\bibitem{bubble-flux-isca13}
H.~Yang, A.~Breslow, J.~Mars, and L.~Tang.
\newblock {Bubble-Flux}: Precise online qos management for increased
  utilization in warehouse scale computers.
\newblock In {\em ISCA}, pages 607--618, 2013.

\bibitem{split-level-io-sosp15}
S.~Yang, T.~Harter, N.~Agrawal, S.~S. Kowsalya, A.~Krishnamurthy,
  S.~Al-Kiswany, R.~T. Kaushik, A.~C. Arpaci-Dusseau, and R.~H. Arpaci-Dusseau.
\newblock Split-level i/o scheduling.
\newblock In {\em SOSP}, pages 474--489, 2015.

\bibitem{spark}
M.~Zaharia, M.~Chowdhury, M.~J. Franklin, S.~Shenker, and I.~Stoica.
\newblock {Spark: Cluster computing with working sets}.
\newblock HotCloud, page~10, Berkeley, CA, USA, 2010.

\bibitem{virtualclock-zhang1989new}
L.~Zhang.
\newblock A new architecture for packet switching network protocols.
\newblock Technical report, MASSACHUSETTS INST OF TECH CAMBRIDGE LAB FOR
  COMPUTER SCIENCE, 1989.

\bibitem{hadoop-scheduling}
W.~Zhang, S.~Rajasekaran, S.~Duan, T.~Wood, and M.~Zhuy.
\newblock Minimizing interference and maximizing progress for hadoop virtual
  machines.
\newblock {\em SIGMETRICS Perform. Eval. Rev.}, 42(4):62--71, 2015.

\bibitem{cpi2-eurosys13}
X.~Zhang, E.~Tune, R.~Hagmann, R.~Jnagal, V.~Gokhale, and J.~Wilkes.
\newblock {CPI$^2$: CPU} performance isolation for shared compute clusters.
\newblock In {\em EuroSys}, pages 379--391, 2013.

\bibitem{justitia-nsdi22}
Y.~Zhang, Y.~Tan, B.~E. Stephens, and M.~Chowdhury.
\newblock {RDMA} performance isolation with justitia.
\newblock In {\em NSDI}, 2022.

\bibitem{prefetch-hardware3}
D.~F. {Zucker}, R.~B. {Lee}, and M.~J. {Flynn}.
\newblock Hardware and software cache prefetching techniques for {MPEG}
  benchmarks.
\newblock {\em IEEE Transactions on Circuits and Systems for Video Technology},
  10(5):782--796, 2000.

\end{thebibliography}
\clearpage
\appendix
\mysection{Extended Motivation \label{sec:extended-motivation}}
\begin{figure} [!ht]
    \centering
    \begin{tabular}{cc}
    \includegraphics[scale=.27]{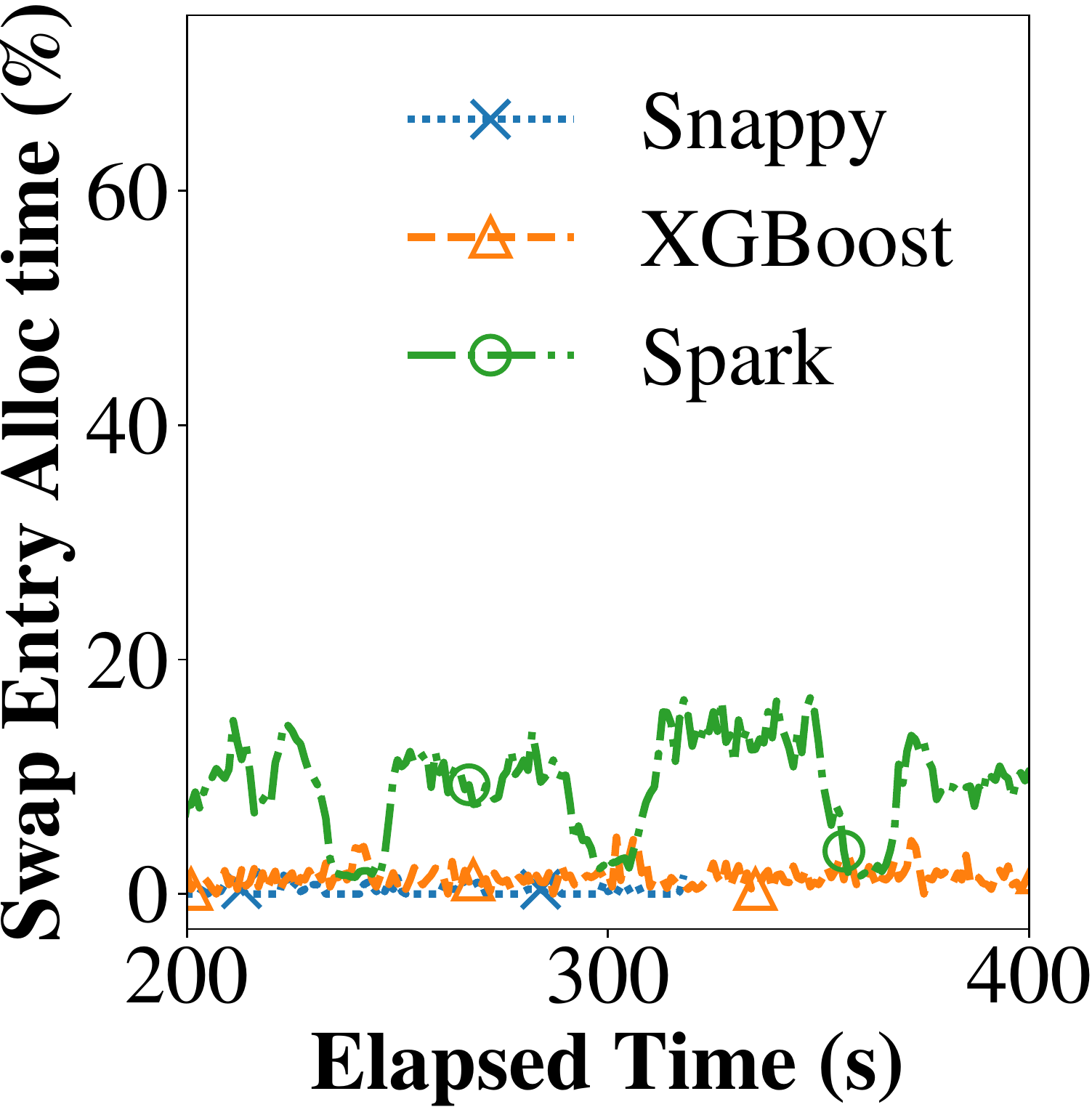} &
    \includegraphics[scale=.27]{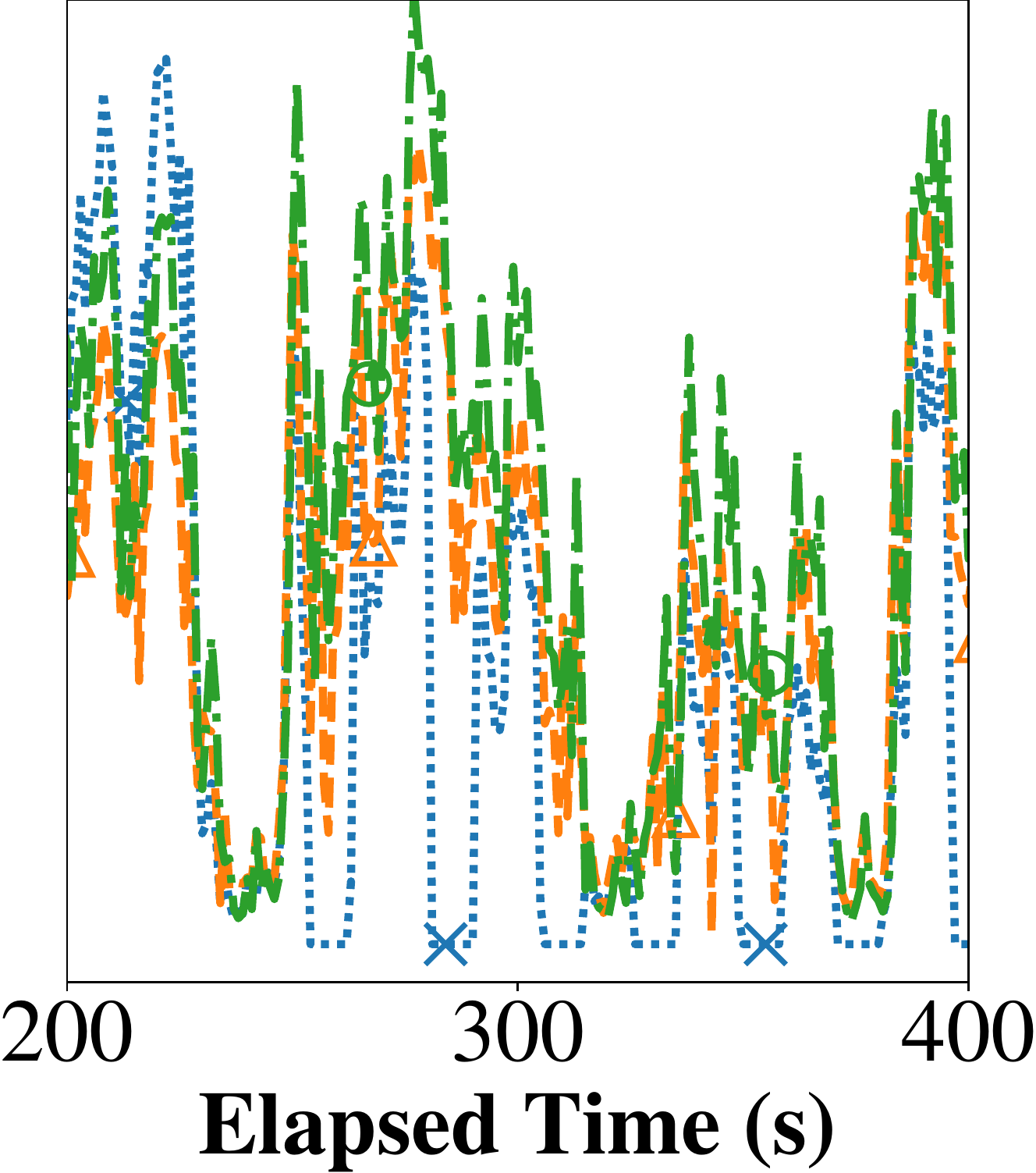}\\
    (a) Running individually. & (b) Corunning.
    \end{tabular}
    \caption{Percentage of time spent on swap entry allocation when applications run individually (a) and together (b).\label{fig:lock-contention}}
\end{figure}

Figure~\ref{fig:lock-contention} compares the percentage of time spent on swap entry allocation between individual runs and co-runs under Linux 5.5. As shown, each application, when co-run with other applications, spend significantly more time on allocating swap entries due to the increased locking time.

\mysection{Recent Kernel Development\label{sec:kernelpatches}}
As an optimization in Linux 5.5, the kernel keeps swap entries for clean pages\textemdash when clean pages are evicted, they do not need to be written back if their swap entries are not released for other allocations. Once a page becomes dirty, its swap entry must be immediately released. Clearly, this approach works for read-intensive applications where most pages are clean, but not for write-intensive workloads such as Spark. We tried various entry-keeping thresholds (\ie, entry keeping starts when the percentage of available swap entries exceeds this threshold) between 25\% and 75\%, and saw only marginal performance differences ($<$5\%) across our programs.

We have closely followed the kernel development since the release of Linux 5.5 and found two recent patches related to our approach. 
These two patches, submitted by Intel and merged into the kernel at 5.8, also attempt to optimize locking overhead at swap entry allocation. The idea of the first patch~\cite{patch1} is using fine-grained locking\textemdash dividing swap entries into \emph{clusters} and assigning each core a random cluster upon an allocation request. The second patch~\cite{patch2} performs batch entry allocation by scanning more swap entries while holding the lock to make each batch larger.  
Note that our adaptive allocation algorithm solves a much bigger problem than these patches\textemdash \tool \emph{avoids} allocating entries for most swap-outs, while these patches reduce the overhead of locking for each allocation. As such, \tool is completely lock-free for reserved entries while these patches must still go through the allocation path, requiring locking if multiple cores are assigned the same cluster (\ie, core collision). 

In fact, the probability of collision increases quickly with the number of cores. As shown below in Figure~\ref{fig:5.14-memcached}, the allocation performance of these patches degrades super-linearly when the number of cores exceeds 24. 
Another major drawback is that none of these patches build on isolated swap partitions.
Lack of swap partition isolation makes applications search for swap entries globally, which can still result in interference\textemdash applications such as Spark can quickly saturate these clusters with all its executor threads, making other applications wait before they can obtain the locks. By reserving entries, our algorithm significantly reduces the number of entry allocation requests (due to entry reusing) and the cost of each allocation (due to reduced lock contention).

\begin{figure} [h!]
    \centering
    \includegraphics[scale=.37]{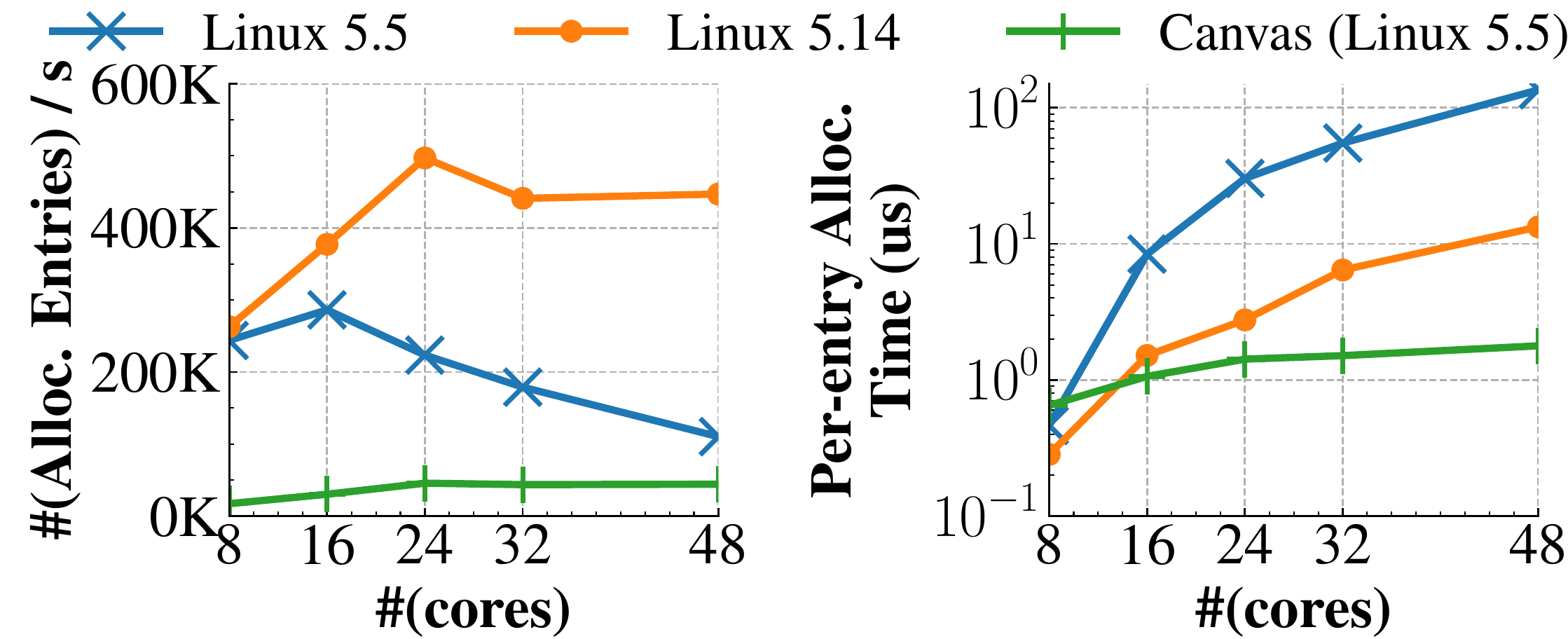} 
    \begin{tabular}{cc}
     (a) Swap entry allocation rate. & (b) Per-entry allocation time.
    \end{tabular}
    \caption{Entry allocation comparison between \tool and the allocation algorithm when Memcached runs on Linux 5.14 on RAMDisk.\label{fig:5.14-memcached}}
\end{figure}

\MyPara{Comparison with Linux 5.5 and Linux 5.14.}
As the kernel is fast evolving and our latest InfiniBand driver is only compatible with Linux 5.5,  we compared the swap-entry allocation performance between \tool, Linux 5.5, and the latest Linux 5.14 over RAMDisk, by running Memcached with varying (8 -- 48) cores.

As Figure~\ref{fig:5.14-memcached}(a) shows, our adaptive entry reservation algorithm 
reduces the allocation rate by several orders of magnitude compared to Linux 5.14. Note that the allocation rate under Linux 5.5 drops as the number cores increases because each allocation takes much longer and hence the swap-out throughput (\ie, allocation throughput) reduces (\ie, the application runs slower).

Figure~\ref{fig:5.14-memcached}(b) compares our algorithm with Linux 5.5 and Linux 5.14 on per-entry allocation time. As shown, the optimization in \cite{patch1,patch2} is unscalable\textemdash as the number of cores increases, the per-entry allocation cost increases significantly. In fact, the allocation cost grows superlinearly after 24 cores due to core collision. On the contrary, \tool's per-entry allocation cost remains low and stable. With 48 cores, our algorithm outperforms Linux 5.14's entry allocator (that uses \cite{patch1,patch2}) by \textbf{13$\times$}.

\end{document}